\documentclass[aps, prl, twocolumn, groupedaddress]{revtex4}%
\usepackage{amsfonts}
\usepackage{amsmath}
\usepackage{amssymb}
\usepackage{graphicx}
\usepackage{bm}
\usepackage{subfigure}
\usepackage{color}
\usepackage[colorlinks=true,linktocpage=true,breaklinks=true]{hyperref}%
\providecommand{\U}[1]{\protect\rule{.1in}{.1in}}
\providecommand{\U}[1]{\protect\rule{.1in}{.1in}}

\begin{document}
\title{Photoninduced Weyl half-metal phase and spin filter effect from topological
Dirac semimetals}
\author{Xiao-Shi Li$^{1}$}
\author{Chen Wang$^{3}$}
\author{Ming-Xun Deng$^{1}$}
\email{dengmingxun@scnu.edu.cn}
\author{Hou-Jian Duan$^{1}$}
\author{Pei-Hao Fu$^{1}$}
\author{Rui-Qiang Wang$^{1}$}
\email{wangruiqiang@m.scnu.edu.cn}
\author{L. Sheng$^{2}$}
\author{D. Y. Xing$^{2}$}
\email{dyxing@nju.edu.cn}
\affiliation{$^{1}$Guangdong Provincial Key Laboratory of Quantum Engineering and Quantum Materials, GPETR Center for Quantum Precision Measurement, SPTE, South China Normal University, Guangzhou 510006, China }
\affiliation{$^{2}$National Laboratory of Solid State Microstructures and Department of Physics, Nanjing University, Nanjing 210093, China}
\affiliation{$^{3}$Lab for Computational Imaging Technology and Engineering, School of Electronic Science and Engineering, Nanjing University, Nanjing 210023, China}
\date{\today }

\begin{abstract}
Recently discovered Dirac semimetals (DSMs) with two Dirac nodes, such as
Na$_{3}$Bi and Cd$_{2}$As$_{3}$, are regarded to carry the $\mathbb{Z}_{2}$
topological charge in addition to the chiral charge. Here, we study the
Floquet phase transition of $\mathbb{Z}_{2}$ topological DSMs subjected to a
beam of circularly polarized light. Due to the resulting interplay of the
chiral and $\mathbb{Z}_{2}$ charges, the Weyl nodes are not only
chirality-dependent but also spin-dependent, which constrains
the behaviors in creation and annihilation of the Weyl nodes in pair.
Interestingly, we find a novel phase: One spinband is in Weyl semimetal phase
while the other spinband is in insulator phase, and we dub it Weyl half-metal
(WHM) phase. We further study the spin-dependent transport in
a Dirac-Weyl semimetal junction and find a spin filter effect as a fingerprint
of existence of the WHM phase. The proposed spin filter effect, based on the
WHM bulk band, is highly tunable in a broad parameter regime and robust
against magnetic disorder, which is expected to overcome the
shortcomings of the previously proposed spin filter based on the topological
edge/surface states. Our results offer a unique opportunity to explore the
potential applications of topological DSMs in spintronics.

\end{abstract}
\maketitle

Great interest is recently triggered in spintronics towards three-dimensional
(3D) topological semimetal materials\cite{RevModPhys.90.015001,Liu864}, such
as Dirac semimetals (DSMs) and Weyl semimetals (WSMs). A DSM is a 3D
counterpart of graphene, in which the conduction and valence bands touch, near
the Fermi surface, at certain discrete Dirac point (DP). The DPs are usually unstable,
because of the strong repulsion between degenerate
bands\cite{Yang:2014aa,PhysRevB.91.121101}. By breaking the time-reversal (TR)
or spatial-inversion (SI) symmetry, a single DP can split into a pair of Weyl
nodes, leading to the phase transition from a DSM to a
WSM\cite{PhysRevX.9.011039,PhysRevB.98.085149,PhysRevB.96.155141}.
Accompanied with this, there
emerge topological surface states, which are protected by the quantized Berry
flux, to connect the two split Weyl nodes.

According to the classification\cite{Yang:2014aa}, there
are two distinct classes of 3D DSMs: One class is the topologically trivial
DSMs possessing a single DP at a time-reversal invariant momentum and the
other is the topologically nontrivial DSMs possessing a pair of DPs created by
band inversion, such as Na$_{3}$Bi\cite{Liu864} and Cd$_{2}$As$_{3}%
$\cite{Neupane:2014aa} compounds. Usually, the dispersion of the nodes is
described by the momentum coupled to pseudospin while the real spin is
suppressed due to spin degeneracy. Nevertheless, in
topological DSMs Na$_{3}$Bi and Cd$_{2}$As$_{3}$, there is well-defined real
spin and two Weyl nodes at each DP are spin-resolved.
Since the Weyl nodes belong to different irreducible
representations, two Weyl nodes at the same DP cannot be coupled in pair and
have to seek for a partner with the same spin from the other DP. As a
consequence, the two DPs including two pairs of Weyl nodes, separated by a net
momentum in the Brillouin zone (BZ), are connected by two spin-polarized Fermi
arcs\cite{A3B_PhysRevB.85.195320,Cd3As2_PhysRevB.88.125427,Liu864,PhysRevB.83.205101,Yang:2014aa}%
. This scenario resembles the topological insulators, carrying a non-trivial
$\mathbb{Z}_{2}$ topological charge, and thus was named as $\mathbb{Z}_{2}$
topological DSMs\cite{PhysRevB.91.121101,PhysRevLett.117.136602}.

To explore the unique properties of DSMs/WSMs from transport
measurements, many
works\cite{Xiong413,Li:2015aa,Zhang:2017aa,PhysRevX.5.031023,PhysRevLett.122.036601,PhysRevX.8.031002}
have been devoted to the longitudinal negative magnetoresistance effect due to
the chiral anomaly. In addition to the chiral anomaly, the $\mathbb{Z}_{2}$
DSMs also carry a $\mathbb{Z}_{2}$ topological charge, exhibiting the
$\mathbb{Z}_{2}$ quantum anomaly\cite{PhysRevLett.117.136602}. A natural
question to ask in this regard is whether the existence of $\mathbb{Z}_{2}$
topological charge also manifests itself in any way in transport. Burkov and
Kim\cite{PhysRevLett.117.136602} addressed that in magnetotransport, the
$\mathbb{Z}_{2}$ topological order manifested as the spin Hall effect can lead
to observable effects by narrowing the dependence of the positive
magnetoconductivity on the angle between the current and the applied magnetic
field, which provides a possible explanation for a recent
experiment\cite{Xiong413}. Besides, the influence of interplay between the
$\mathbb{Z}_{2}$ and chiral anomalies on magnetoconductivity has been studied
in a relativistic hydrodynamics limit\cite{JHEP2018}.

Instead of the magnetotransport, the main objective of our work is to exploit
the joint influence of $\mathbb{Z}_{2}$ topological and chiral charges on spin
transport. In this Letter, we apply a beam of circularly
polarized light (CPL), which is widely adopted to induce topological phase
transitions of
matter\cite{PhysRevX.3.031005,PhysRevLett.114.056801,PhysRevLett.115.106403,PhysRevLett.116.026805,PhysRevLett.117.087402,PhysRevB.95.115102,FU20173499}%
, to drive the phase transition of $\mathbb{Z}_{2}$ DSMs. We find that the CPL
can create first and then annihilate the Weyl nodes in pair and finally gap
them out, during which the node pairs associated with different spin
orientations exhibit different response to the CPL. Consequently, the
$\mathbb{Z}_{2}$ DSM can be driven into a Weyl half-metal (WHM) phase. Based
on this, we further analyze the spin-dependent transport in a
Dirac-Weyl semimetal junction and find that the existence of WHM phase is
manifested as a fully spin-polarized current, i.e., the spin filter effect. The half-metallicity was recently
employed to realize the fully spin polarized Weyl loops or the magnetic topological semimetal states\cite{Chang:2016aa,PhysRevB.99.075131}.

\begin{figure}[ptb]
\centering \includegraphics[width=\linewidth]{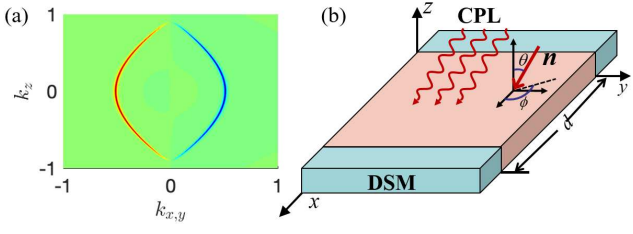}\caption{(a)
Spin-resolved Dirac points connected by two Fermi arcs projected on $k_{z}%
$-$k_{x,y}$ plane in a $\mathbb{Z}_{2}$ topological DSMs. The filled colors
denote the magnitude of the averaged surface spin density of states
$\langle\sigma_{z}\rangle$, with red, green and blue for $\langle\sigma
_{z}\rangle\{>,=,<\}0$, respectively. (b) Schematic illustration of the
DSM-WSM tunnel junction, in which the middle region ($0<x<d$) is irradiated by
a beam of CPL along direction $\mathbf{n}=(\sin\theta\cos\phi,\sin\theta
\sin\phi,\cos\theta)$, where $\theta$ and $\phi$ are the polar and azimuthal
angles in the spherical coordinate system.}%
\label{figDSM}%
\end{figure}

\emph{Floquet effective Hamiltonian-} Let us start from a four-band effective
Hamiltonian
\begin{align}
H(\mathbf{k})  &  =\epsilon_{0}(\mathbf{k})+M(\mathbf{k})\tau_{z}+v_{0}%
(k_{x}\sigma_{z}\tau_{x}-k_{y}\tau_{y})\nonumber\\
&  +\frac{B_{3}k_{z}}{2}(k_{+}^{2}\sigma_{-}+k_{-}^{2}\sigma_{+})\tau_{x},
\label{eq_Hal}%
\end{align}
with $k_{\pm}=k_{x}\pm ik_{y}$ and $\sigma_{\pm}=\sigma_{x}\pm i\sigma_{y}$,
which is widely adopted to describe topological properties of $\mathbb{Z}_{2}$
topological DSMs \textrm{Na}$_{3}$\textrm{Bi}\cite{A3B_PhysRevB.85.195320}
and\textrm{ Cd}$_{3}\mathrm{As}_{2}$\cite{Cd3As2_PhysRevB.88.125427}. Here,
$\sigma_{x,y,z}$ ($\tau_{x,y,z}$) are Pauli matrices for the electron spin
(orbital parity or pseudospin), $\epsilon_{0}(\mathbf{k})=C_{0}+C_{1}k_{z}%
^{2}+C_{2}(k_{x}^{2}+k_{y}^{2})$ and $M(\mathbf{k})=M_{0}-M_{1}k_{z}^{2}%
-M_{2}(k_{x}^{2}+k_{y}^{2})$, with $C_{j}$, $M_{j}$, $B_{3}$, and $v_{0}$ as
material parameters. If we only concentrate on the neighborhood of each
gap-crossing point and neglect the high-order terms $B_{3}$, Hamiltonian
(\ref{eq_Hal}) describes two superimposed copies of WSMs, related respectively
to two spin species which is a conserved quantity. It is easy to find that,
for $M_{0}/M_{1}>0$, there exist two DPs at $\mathbf{k}_{c}=(0,0,\pm
\sqrt{M_{0}/M_{1}})$, each of which contains two spin-resolved Weyl nodes. It is
distinct from an ordinary DSM that the two Weyl nodes at the same DP here are
nonpaired since two Weyl sectors are protected from mixing by the
$\mathbb{Z}_{2}$ symmetry. The $\mathbb{Z}_{2}$ DSM structure is confirmed by
the existence of two surface Fermi arcs, as shown in Fig. \ref{figDSM}(a),
connecting the two paired Weyl nodes from different DPs. The spin texture of
the surface states has a helical structure, resembling the surface states
of topological insulators.

When a beam of CPL irradiates along the direction $\mathbf{n}=(\sin\theta
\cos\phi,\sin\theta\sin\phi,\cos\theta)$ as illustrated in Fig. \ref{figDSM}%
(b), where $\theta$ and $\phi$ are the polar and azimuthal angles in the
spherical coordinate system, the Dirac fermions can be described by a
time-dependent Hamiltonian $\mathcal{H}(\mathbf{k},t)=H(\mathbf{k}%
+e\mathbf{A}/\hbar)$, where
\begin{equation}
\mathbf{A}=A_{0}[\cos(\omega t)\mathbf{e}_{1}-\eta\sin(\omega t)\mathbf{e}%
_{2}]
\end{equation}
is vector potential for the CPL with amplitude $A_{0}$, frequency $\omega$ and
$\eta=\pm1$ for right/left polarization. The spacial dependence of $A_{0}$ can be neglected, as we consider that the light wavelength is much larger than the device\cite{supplement}. Here, $\mathbf{e}_{1}=(\cos\theta
\cos\phi,\cos\theta\sin\phi,-\sin\theta)$\ and $\mathbf{e}_{2}=(\sin\phi
,-\cos\phi,0)$, satisfying $\mathbf{e}_{1}\cdot\mathbf{e}_{2}=0$, are two unit
vectors perpendicular to the incident direction of the light. Employing the
Floquet
theorem\cite{PhysRevX.3.031005,PhysRevLett.114.056801,PhysRevLett.115.106403,PhysRevLett.116.026805,PhysRevLett.117.087402,PhysRevB.95.115102,FU20173499}
and focusing ourselves on the off-resonant regime, in which $\hbar\omega$ is
greater than the width of the static energy band, we derive an effective
Hamiltanion\cite{supplement}
\begin{align}
&  \mathcal{H}_{\mathrm{eff}}(\mathbf{k}) =\tilde{\epsilon}_{0}%
(\mathbf{k})+\tilde{M}(\mathbf{k})\tau_{z}+v_{0}(k_{x}\sigma_{z}\tau
_{x}-k_{y}\tau_{y})-\lambda\sigma_{z}\tau_{z}\nonumber\\
&  +v_{1}k_{z}(\cos\phi\tau_{x}-\sin\phi\sigma_{z}\tau_{y})-v_{2}(k_{x}%
\tau_{x}-k_{y}\sigma_{z}\tau_{y}), \label{eq_HOR}%
\end{align}
where $B_{3}$ is dropped due to smallness, and denote $\lambda=\eta\frac
{v_{0}^{2}k_{A}^{2}\cos\theta}{\omega}$, $v_{1}=\eta\frac{2M_{1}k_{A}^{2}%
\sin\theta}{\omega}v_{0}$, and $v_{2}=\eta\frac{2M_{2}k_{A}^{2}\cos\theta
}{\omega}v_{0}$, with $k_{A}=eA_{0}/\hbar$ characterizing the CPL intensity.
The renormalized parameters in the first two terms are given by
$\tilde{\epsilon}_{0}(\mathbf{k})=\epsilon_{0}(\mathbf{k})|_{C_{0}%
\rightarrow C_{0}+C_{0}^{\prime}}$ and $\tilde{M}(\mathbf{k}%
)=M(\mathbf{k})|_{M_{0}\rightarrow M_{0}-M_{0}^{\prime}}$, where
$C_{0}^{\prime}=k_{A}^{2}(\frac{1+\eta^{2}}{2}C_{2}+C_{-}\sin^{2}\theta)$ and
$M_{0}^{\prime}=k_{A}^{2}(\frac{1+\eta^{2}}{2}M_{2}+M_{-}\sin^{2}\theta)$,
with $C_{-}=(C_{1}-C_{2})/2$ and $M_{-}=(M_{1}-M_{2})/2$.

\begin{figure}[ptb]
\centering \includegraphics[width=\linewidth]{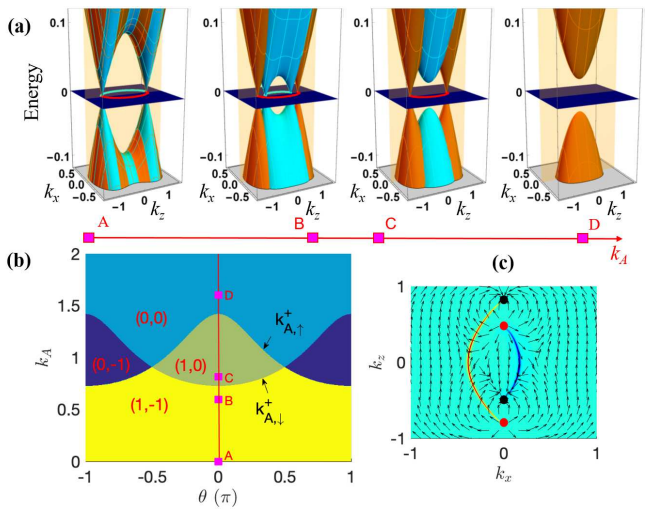} \caption{(a)
Evolution of the spin-resolved quasienergy spectrum with the CPL intensity
$k_{A}=(0,0.6,0.8,1.6)\times\text{nm}^{-1}$ from left to the right side, which
correspond to points A-D marked in (b). The other CPL parameters are $\hbar\omega=1.0$ \textrm{eV}\cite{Sentef:2015aa,Sie2018}, $\phi=0$ and $\theta=0$. The brown and blue
regions represent the spin-$\uparrow$ and spin-$\downarrow$ Weyl sectors,
respectively, and the lines connecting the nodes denote the surface Fermi
arcs. (b) The phase diagram characterized with the spin-dependent Chern fluxes
($C_{\uparrow},C_{\downarrow}$) in the $k_{A}$-$\theta$ parameter space, and
(c) the Berry curvature distribution corresponding to point B, where the red
and black filled circles stand for the Weyl nodes with a positive and negative
chiral charge, respectively. The calculation is carried out for Na$_{3}$Bi,
whose parameters can be found in Ref. \cite{A3B_PhysRevB.85.195320}. }%
\label{figARC}%
\end{figure}

\emph{CPL-driven WHM phase-} As it shows, in the presence of the CPL, the TR
symmetry is broken explicitly, such that the DPs are expected to split into
two pairs of spin-resolved Weyl nodes. The quasienergy spectrum of the
effective Hamiltonian (\ref{eq_HOR}) is%
\begin{equation}
\epsilon_{s,\pm}(\mathbf{k})=\tilde{\epsilon}_{0}(\mathbf{k})\pm
\sqrt{|\tilde{M}_{s}(\mathbf{k})|^{2}+v_{s}^{2}(\tilde{k}_{x,s}%
^{2}+\tilde{k}_{y,s}^{2})}, \label{eq_disper1}%
\end{equation}
where $\tilde{k}_{x,s}=k_{x}+s\frac{v_{1}}{v_{s}}k_{z}\cos\phi$ and
$\tilde{k}_{y,s}=k_{y}+s\frac{v_{1}}{v_{s}}k_{z}\sin\phi$, with
$v_{s}=v_{0}-sv_{2}$ and $\tilde{M}_{s}(\mathbf{k})=\tilde{M}%
(\mathbf{k})-s\lambda$. In contrast to the static spectrum of Hamiltonian
(\ref{eq_Hal}), the CPL-driven spectrum becomes spin-dependent. The evolution
of the spin-resolved quasienergy spectrum with the CPL intensity is plotted in
Fig. \ref{figARC}(a). There are four Weyl nodes located at
\begin{equation}
\mathbf{k}_{c,\pm}^{s}=\pm(sv_{1}\cos\phi,sv_{1}\sin\phi,-v_{s})k_{w,s}/v_{s}
\label{eq_lw}%
\end{equation}
with $k_{w,s}=\sqrt{\frac{M_{0}-(M_{0}^{\prime}+s\lambda)}{M_{1}+v_{1}%
^{2}M_{2}/v_{s}^{2}}}$, which correspond to the vertexes of the dispersion. In
the presence of the CPL, the two Weyl nodes overlapping at one DP are
separated and, simultaneously, the Berry curvature emerges in the whole
momentum space, starting from a Weyl node with positive chiral charge and
ending at one with negative chiral charge [seeing Fig. \ref{figARC}(c)].
Notice that the split Weyl nodes are not only
chirality-dependent but also spin-dependent, as shown in Fig.
\ref{figARC}(a), where the brown and blue regions represent the spin-$\uparrow
$ and spin-$\downarrow$, respectively, and two nodes with the same spin are
connected by an open Fermi arc. Importantly, with increase of $k_{A}$, the
distance $2k_{w,s}$ between the paired Weyl nodes would reduce because of the
CPL-renormalized term $M_{0}^{\prime}$. Consequently, in certain crucial
value, the paired Weyl nodes with opposite chirality can be
merged in momentum space and then gapped out. Notice that the different spin
species have different critical values, which originate from the renormalized
spin-dependent Dirac mass term $s\lambda$. It is interesting to find a phase
as illustrated by the third diagram of Fig. \ref{figARC}(a), where the
spin-$\uparrow$ species still remains at the WSM phase but the
spin-$\downarrow$ species enters the insulator phase, which exhibits a typical
characteristic of well-known spin half-metals. This novel Weyl phase has not
been reported before and we dub it Weyl half-metal (WHM) phase. As $k_{A}$
further increases, both pairs of the Wely nodes could be gapped out, and
finally the system undergoes a transition to an insulating phase, as shown in
the last diagram of Fig. \ref{figARC}(a).

In order to identify these different topological phases, we can further
classify them according to the spin-dependent Chern fluxes ($C_{\uparrow
},C_{\downarrow}$) at the $k_{z}=0$ BZ cross section\cite{Yang:2014aa}. With
the Berry curvature\cite{PhysRevB.91.121101} $\Omega_{s,ij}(\mathbf{k}%
)=\frac{1}{2|d_{s}|^{3}}\epsilon_{abc}d_{s,a}\partial_{i}d_{s,b}\partial
_{j}d_{s,c}$, where $\epsilon_{abc}$ denotes the antisymmetric tensor and
$d_{s}=\left(  sv_{0}\tilde{k}_{x,s},v_{0}\tilde{k}_{y,s}%
,\tilde{M}_{s}(\mathbf{k})\right)  $, we find
\begin{equation}
C_{s}=-\frac{s}{2}[\mathrm{sgn}(M_{2})+\mathrm{sgn}(M_{0}-M_{0}^{\prime
}-s\lambda)]
\end{equation}
for $s$ spin component. In Fig. \ref{figARC}(b), we plot the phase diagram in
the $k_{A}$-$\theta$ parameter space. From the bottom to the top, e.g., points A to D,
the phase is in order DSM phase ($k_{A}=0$), WSM phase ($1,-1$), WHM phase
($1,0$) or ($0,-1$), and normal insulating phase ($0,0$), whose dispersions
are depicted in Fig. \ref{figARC}(a), respectively. The phase boundaries are
determined by the equation $M_{2}+M_{-}\sin^{2}\theta+s\eta v_{0}^{2}%
\cos\theta/\omega=M_{0}/k_{A}^{2}$ via $k_{w,s}=0$. Subsequently, we can
derive the crucial values for the light intensity to be
\begin{equation}
k_{A,s}^{\eta}=\sqrt{\frac{M_{0}}{M_{2}+M_{-}\sin^{2}\theta+s\eta v_{0}%
^{2}\cos\theta/\omega}}.\label{eq_As}%
\end{equation}
The WHM phase emerges in the range of ($k_{A,\downarrow}^{+},k_{A,\uparrow
}^{+}$) for $|\theta|<\frac{\pi}{2}$, where only the spin-$\uparrow$ WHM phase
exists. The spin-$\downarrow$ WHM phase can be realized just by reversing the
polarization $\eta$ of the CPL or by tuning the incident direction
$|\theta|>\frac{\pi}{2}$ to be in the WHM region ($0,-1$). If $k_{A}%
>k_{A,\uparrow}^{+}$, the CPL-driven DSM would cross over to a normal
insulator with spin-dependent Chern fluxes ($0,0$), where both pairs of the
Weyl nodes are gapped out. Also, we can tune the thresholds $k_{A,\downarrow
}^{+}$ and $k_{A,\uparrow}^{+}$ by changing the incident direction of the CPL
due to the rotation of the Weyl nodes in the $k_{x}$-$k_{z}$ plane. For a
specific value $\theta=\pm\pi/2$, the thresholds $k_{A,\downarrow}%
^{+}=k_{A,\uparrow}^{+}$ and thus the WHM phase region vanishes.
\begin{figure}[ptb]
\centering \includegraphics[width=\linewidth]{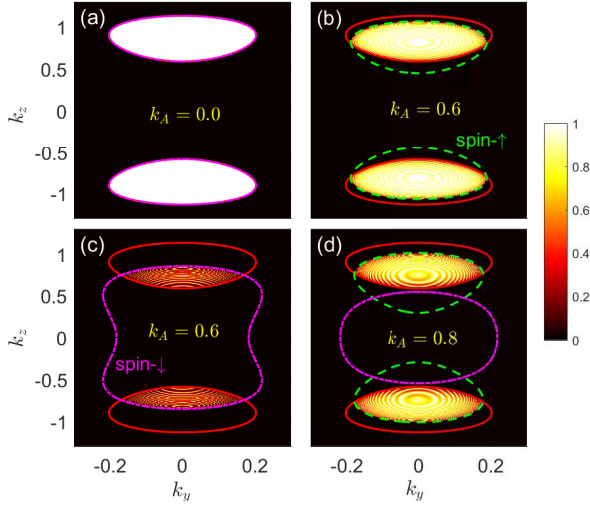}
\caption{Distribution of the spin-dependent transmission probability
$T_{\uparrow}(k_{y},k_{z})$ and $T_{\downarrow}(k_{y},k_{z})$ on the $k_{y}%
$-$k_{z}$ plane, with (a) for $k_{A}=0$ where $T_{\uparrow}(k_{y},k_{z})$
coincides with $T_{\downarrow}(k_{y},k_{z})$ completely, (b)-(c) for
$k_{A}=0.6$, and (d) for $k_{A}=0.8$ where $T_{\downarrow}(k_{y},k_{z})=0$ .
The lines represent the projections of the Fermi surfaces, with red-solid,
green-dash and pink-dash-dot for the reservoirs, the spin-$\uparrow$ and
spin-$\downarrow$ components in the middle WSM region, respectively. Here,
$E_{\mathrm{F}}=0.05$ \textrm{eV},
$d=400$\textrm{nm}, $\phi=\pi/2$, and other parameters are the
same as Fig. \ref{figARC}(a).}%
\label{figSUM}%
\end{figure}

\emph{CPL-modulated spin-dependent transmission-} In the following, we study
the transport fingerprint related to the WHM phase. We construct a DSM/WSM
sandwich junction, as shown in Fig. \ref{figDSM}(b), where the middle WSM
region $0<x<d$ is formed due to the irradiation by a beam of CPL. The
eigenequation for the DSM reservoirs can be obtained as%
\begin{equation}
\lbrack E-\epsilon_{0}(\mathbf{k})]^{2}=[M(\mathbf{k})]^{2}+v_{0}^{2}%
(k_{x}^{2}+k_{y}^{2}), \label{eq_root}%
\end{equation}
from which we can determine $k_{x}$ for fixed $k_{y}$, $k_{z}$ and the Fermi
energy $E=E_{\mathrm{F}}$. We are interested in the electron
transport near the gap-closing points, and so the unimportant
terms $\epsilon_{0}(\mathbf{k})$ and $M_{2}(k_{x}^{2}+k_{y}^{2})$ can be
neglected. It is easy to find the wavefuncitons in different regions. By
matching the wavefunctions at the interfaces $\psi_{s}(x=0^{-})=\psi
_{s}(x=0^{+})$ and $\psi_{s}(x=d^{-})=\psi_{s}(x=d^{+})$, we can obtain the
transmission coefficient as
\begin{equation}
t_{s}=\frac{2\mathcal{F}_{1}\mathcal{F}_{2}e^{-i(k_{1}+s\frac{v_{1}}{v_{s}%
}k_{z}\cos\phi)d}}{2\mathcal{F}_{1}\mathcal{F}_{2}\cos(k_{2}d)-i(\mathcal{F}%
_{1}^{2}+\mathcal{F}_{2}^{2}+\Delta_{y}^{2})\sin(k_{2}d)}, \label{eq_ts}%
\end{equation}
where $k_{1}$ represents the positive root of Eq. (\ref{eq_root}) and $k_{2}$
can be obtained similarly from the eigenequation of the CPL-driven region. In
Eq. (\ref{eq_ts}), $\mathcal{F}_{1}=v_{0}k_{1}[E-\tilde{M}_{s}%
(\mathbf{k})]$, $\mathcal{F}_{2}=v_{s}k_{2}[E-M(\mathbf{k})]$ and%
\begin{equation}
\Delta_{y}=v_{0}k_{y}[E-\tilde{M}_{s}(\mathbf{k})]-v_{s}\tilde{k}%
_{y,s}[E-M(\mathbf{k})].
\end{equation}
The spin-dependent transmission probability can be obtained by $T_{s}%
(k_{y},k_{z})=\left\vert t_{s}\right\vert ^{2}$. For $d\gg1$, the transmission
probability is finite only when $k_{1,2}\in\operatorname{real}$, i.e., $k_{y}$
and $k_{z}$ are within the overlapping region of the projected Fermi surfaces
of the reservoirs and the CPL-driven WSM.

The distributions of the spin-dependent transmission probability $T_{\uparrow
}(k_{y},k_{z})$ and $T_{\downarrow}(k_{y},k_{z})$ on the $k_{y}$-$k_{z}$ plane
are plotted in Fig. \ref{figSUM}. In the absence of the CPL, $T_{\uparrow
}(k_{y},k_{z})=T_{\downarrow}(k_{y},k_{z})=1$ for all the incident electron
states, as shown in Fig. \ref{figSUM}(a), because of the perfect matching of
the Fermi surfaces between different regions. With introduction of the CPL, the Weyl nodes are split and move in the $k_{z}$ axis, which leads to mismatch of wave vectors between the middle region and
the reservoirs\cite{PhysRevLett.102.026807,Young:2009aa}. Due to the broken TR symmetry, two spin species exhibit different
splitting distances along $k_{z}$ and so different shapes of the Fermi
surfaces in Figs. \ref{figSUM}(b) and (c). Consequently, the spin-$\uparrow$ Fermi surface in the middle WSM
region has a larger overlapping with that in the DSM reservoirs (denoted by
the red solid lines) whereas the spin-$\downarrow$ Fermi surface has smaller
overlapping. Naturally, there emerges interesting spin-dependent electron
transport processes. With further increase of irradiation intensity $k_{A}$,
the spin-$\downarrow$ Fermi surfaces in the middle region would disconnect
from that of the reservoirs and so has no contribution to the electron
transport, as shown in Fig. \ref{figSUM}(d). In this
situation, there will emerge a fully spin-polarized current, implying that the
irradiated region has entered the WHM phase ($1,0$).

\begin{figure}[ptb]
\centering\includegraphics[width=\linewidth]{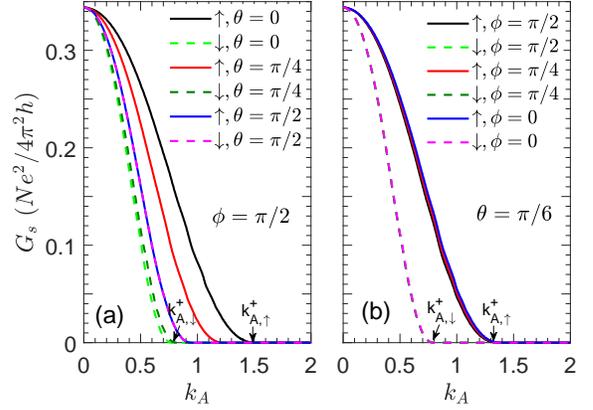}\caption{
Spin-dependent conductances $G_{\uparrow}$ and $G_{\downarrow}$ as a function
of the CPL intensity $k_{A}$ for (a) different $\theta$ with $\phi=\pi/2$ and for
(b) different $\phi$ with $\theta=\pi/6$. Other parameters are the same as in
Fig. \ref{figSUM}.}%
\label{figGSP}%
\end{figure}\emph{Fully spin-polarized conductance-} The spin-dependent
conductance can be calculated by the Landauer formula
\begin{equation}
G_{s}=\frac{e^{2}}{h}\sum\limits_{k_{y},k_{z}}T_{s}(k_{y},k_{z}),
\end{equation}
where the summation runs over all the incident modes at the Fermi surface,
i.e., the $k_{y}$-$k_{z}$ surface BZ. In Fig. \ref{figGSP}(a), we plot the
spin-dependent conductance $G_{\uparrow}$ and $G_{\downarrow}$ as a function
of the CPL intensity $k_{A}$ for different polar angles $\theta$. As $k_{A}$
increases, the spin-dependent conductances $G_{\uparrow}$ and $G_{\downarrow}$
will decrease due to suppression of the transmission probability, and after
certain thresholds they vanish completely. Although both conductances reduce
with $k_{A}$, $G_{\downarrow}$ decays with a faster rate than $G_{\uparrow}$,
such that the threshold $k_{A,\downarrow}^{+}$ for spin-$\downarrow$ current
is lower than $k_{A,\uparrow}^{+}$ for spin-$\uparrow$ current. For $k_{A}%
\in(k_{A,\downarrow}^{+},k_{A,\uparrow}^{+})$, the electric current is
spin-$\uparrow$ fully-polarized, exhibiting a perfect spin filter effect. The
spin filter region ($k_{A,\downarrow}^{+},k_{A,\uparrow}^{+}$) is quite
sensitive to the incident direction $\theta$ of the CPL. As the polar angle
deviates from $\theta=0$, the spin filter region ($k_{A,\downarrow}%
^{+},k_{A,\uparrow}^{+}$), as shown in Fig. \ref{figGSP}(a), will reduce. For
a specific value $\theta=\pi/2$, there is no spin-polarized conductance
because the WHM phase does not exits in this situation, as seen in Fig.
\ref{figARC}(b). In contrast, the spin filter region, as shown in Fig.
\ref{figGSP}(b), is almost independent on the azimuthal angle $\phi$ of the
CPL, since the thresholds given by Eq. (\ref{eq_As}) is irrelevant with $\phi
$.

\emph{Conclusion and remarks-} We have theoretically studied the Floquet phase
transition of $\mathbb{Z}_{2}$ topological DSMs subjected to a beam of CPL.
Due to the interplay of the chiral and $\mathbb{Z}_{2}$ charges, the Weyl
nodes are not only chirality-dependent but also
spin-dependent, which constrains the behaviors in creation and annihilation of
the Weyl nodes in pair. During the evolution of Weyl nodes with the CPL
intensity, we find an entirely new WHM phase: One spinband is in WSM phase while
the other spinband is in insulator phase, which simultaneously possesses the
characteristic of both WSMs and half-metals. This novel phase has not been
reported before. We have also checked that an applied magnetic field/magnetic
perturbation or a linearly polarized light can not achieve this WHM phase,
because they cannot realize the creation and annihilation of the Weyl nodes
pair at the same time\cite{supplement}.

We further study the spin-dependent transport and find a spin
filter effect as a fingerprint of existence of the WHM phase. Indeed, spin filter
transistor is an important device in spintronics, with numerous works focus on its implementation
using the topological edge/surface
states\cite{Rycerz:2007aa,Brune:2012aa,PhysRevLett.105.057202,PhysRevB.89.195303,PhysRevApplied.10.034059}%
. However, these have several typical shortcomings: (1) Disturbance from unpolarized
bulk band; (2) Easily suffering from magnetic disorder; (3)
Only appearing at several specific energy position, which is difficult to
manipulate experimentally. Recently, Tasi $et.$$al.$\cite{Tsai:2013aa}
proposed a silicene-based spin filter device to reduce above shortcomings by
taking advantage of the bulk carriers. Here, we propose the spin filter effect
based on the WHM bulk band, which is highly tunable in a broad parameter
regime and robust against magnetic disorder. Moreover, we can
implement it, no needing other additional conditions in contrast to previous works.

In realistic experiments, one can realize our proposed model using a beam of gaussian profile light or replacing Weyl-semimetal reservoirs with normal-metal electrodes, as discussed in supplementary material\cite{supplement}. The threshold of the incident
radiation intensity for the system entering the WHM phase is determined by Eq.(7).
For Na$_{3}$Bi\cite{A3B_PhysRevB.85.195320}, with $k_{A}=eA_{0}/\hbar$ and $A_{0}=E_{0}/\omega$, we estimate the
threshold of the irradiated electric field to be
$E_{0}\simeq7.2\times10^{8}V/m$ for $\hbar\omega=1$ $\mathrm{eV}$. For Cd$_{3}$As$_{2}$\cite{PhysRevB.95.174505}, the threshold can be reduced by an
order in magnitude $E_{0}\simeq5\times10^{7}V/m$ for $\hbar\omega=0.5$$\mathrm{eV}$. These radiation intensities are within the current experimental accessibility\cite{science3424532013,NaturePhysics123062016}. In the normal metal/Weyl semimetal/normal metal junction, more lower critical radiation intensity can be achieved.

This work was supported by the National Natural Science Foundation of China
under Grants No. 11874016, \textcolor{red}{11904107}, the Key Program for Guangdong NSF of China under
Grant No. 2017B030311003, GDUPS(2017), the State Key Program for
Basic Researches of China under Grants No. 2015CB921202 (L. S.) and No. 2017YFA0303203 (D. Y. X.), and by the Innovation Project
of Graduate School of South China Normal University.

Xiao-Shi Li and Chen Wang contributed equally to this work.

\bibliographystyle{apsrev4-1}
\bibliography{biblxsPRL}

\end{document}


\setcounter{equation}{0} \renewcommand{\theequation}{A\arabic{equation}}

\begin{center}
\textbf{Supplementary material A: Derivation for the effective Hamiltonian Eq.
(3)}
\end{center}

In this appendix, we provide the details for derivation of Eq. (3) in the main text.

The Dirac fermions driven by a beam of circularly polarized light (CPL) can be
described by a time-dependent Hamiltonian $\mathcal{H}(\mathbf{k}%
,t)=H(\mathbf{k}+e\mathbf{A}/\hbar)$, where the vector potential $\mathbf{A}$
is given by Eq. (2) of the main text. According to Ref. [9] of the main text,
we give out the system parameters of \textrm{Na}$_{3}$\textrm{Bi} for
convenience as follows: $v_{0}=2.4598$ $eV\mathring{A}$, $C_{0}=-0.06382$
$eV$, $C_{1}=8.7536$ $eV\mathring{A}^{2}$, $C_{2}=-8.4008$ $eV\mathring{A}^{2}$, $M_{0}=-0.08686$ $eV$, $M_{1}=-10.6424$ $eV\mathring{A}^{2}$ and
$M_{2}=-10.3610$ $eV\mathring{A}^{2}$.
Employing the Floquet
theorem\cite{PhysRevX.3.031005,PhysRevLett.114.056801,PhysRevLett.115.106403,PhysRevLett.116.026805,PhysRevLett.117.087402}%
, one can expand $\mathcal{H}(\mathbf{k},t)$ in the Floquet basis as%
\begin{equation}
\mathcal{H}_{n,n^{\prime}}=n\omega\delta_{n,n^{\prime}}+\frac{1}{T}\int%
_{0}^{T}dte^{i(n-n^{\prime})\omega t}\mathcal{H}(\mathbf{k},t)\text{ },
\label{eq_Flo}%
\end{equation}
where $T=2\pi/\omega$ is the period of the CPL. By diagonalizing the Floquet
Hamiltonian $\mathcal{H}$, it is easy to obtain the Floquet states and the
quasienergy spectrum. The Floquet states, in fact, are uniquely and completely
parametrized by quasienergies in the \textquotedblleft first quasienergy
Brillouin zone\textquotedblright\ $-\hbar\omega/2\leqslant\epsilon<\hbar
\omega/2$, where $\epsilon$ denotes the quasienergy.

The retarded Green's function is defined as $G^{r}(\epsilon)=\frac{1}%
{\epsilon+i0^{+}-\mathcal{H}}$, from which one can express the Floquet
retarded Green's function for the $n$-th Floquet band
as\cite{PhysRevB.95.115102}%
\begin{equation}
G_{n,n}^{r}(\epsilon)=\frac{1}{\epsilon^{+}-\mathcal{H}_{n,n}-\sum_{m\neq
n}\mathcal{H}_{n,m}G_{m,m}^{r}(\epsilon)\mathcal{H}_{m,n}}\text{ },
\end{equation}
where we have adopted abbreviation $\epsilon^{+}=\epsilon+i0^{+}$. By
inspecting the poles of the retarded Green's function, we can obtain an
effective Hamiltonian for the $n$-th Floquet band
\begin{equation}
\mathcal{H}_{\mathrm{eff},n}(\mathbf{k})=\mathcal{H}_{n,n}+\sum_{m\neq
n}\mathcal{H}_{n,m}G_{m,m}^{r}(\epsilon)\mathcal{H}_{m,n}\text{ }.
\label{eq_Heff}%
\end{equation}

For a relatively weak radiation field, the properties of the CPL-driven DSM
are dominated by the $|n=0\rangle$ Floquet state, such that we would focus
ourselves on the behavior of this Floquet state. In the limit of off-resonant
approximation, i.e., high-frequency light with $\hbar\omega$ greater than the
width of the static energy band, real photon-absorption processes are
suppressed. In this case, we can safely approximate $G_{n,n}^{r}%
(\epsilon)\simeq-\frac{1}{n\omega}+\mathcal{O}(\frac{1}{\omega^{2}})$ for
$n\neq0$. As a consequence, for the $|n=0\rangle$ Floquet state,
$\mathcal{H}_{\mathrm{eff},n}(\mathbf{k})$ can be closed as%
\begin{equation}
\mathcal{H}_{\mathrm{eff}}(\mathbf{k})=\mathcal{H}_{n,n}\delta_{n,0}%
+\sum_{m>0}\frac{1}{m\omega}[\mathcal{H}_{m,0},\mathcal{H}_{0,m}%
]+\mathcal{O}(\frac{1}{\omega^{2}})\text{ }, \label{eq_Heff1}%
\end{equation}
where we noted $\mathcal{H}_{\mathrm{eff}}(\mathbf{k})\equiv\mathcal{H}%
_{\mathrm{eff},0}(\mathbf{k})$ for brevity. As can be seen, Eq.
(\ref{eq_Heff1}) exactly recovers the result given in Ref.
\cite{PhysRevLett.117.087402}.

By using Hamiltonian (1) and Eq. (2) of the main text, it is easy to verify
$[\mathcal{H}_{n,0},\mathcal{H}_{0,n}]=0$ for $n\geqslant2$. If neglecting the
higher-order terms tied to $B_{3}$, we find $[\sigma_{z},H(\mathbf{k})]=0$ and
the Hamiltonian can be separated into two $2\times2$ blocks, which can be
labelled by the eigenvalues of $\sigma_{z}$, $s=\pm1$. Subsequently, we
obtain
\begin{align}
\mathcal{H}_{0,0}^{s}  &  =\epsilon_{0}(\mathbf{k})+k_{A}^{2}(\frac{1+\eta
^{2}}{2}C_{2}+\frac{C_{1}-C_{2}}{2}\sin^{2}\theta)+V_{0}\left(
\begin{array}
[c]{cc}%
1 & 0\\
0 & -1
\end{array}
\right)  \text{ },\nonumber\\
\mathcal{H}_{1,0}^{s}  &  =\frac{v_{0}k_{A}}{2}\left(
\begin{array}
[c]{cc}%
0 & e^{is\phi}(s\cos\theta-\eta)\\
e^{-is\phi}(s\cos\theta+\eta) & 0
\end{array}
\right)  +V_{1}\left(
\begin{array}
[c]{cc}%
1 & 0\\
0 & -1
\end{array}
\right)  \text{ },\nonumber\\
\mathcal{H}_{0,1}^{s}  &  =\frac{v_{0}k_{A}}{2}\left(
\begin{array}
[c]{cc}%
0 & e^{is\phi}(s\cos\theta+\eta)\\
e^{-is\phi}(s\cos\theta-\eta) & 0
\end{array}
\right)  +V_{-1}\left(
\begin{array}
[c]{cc}%
1 & 0\\
0 & -1
\end{array}
\right)  \text{ }, \label{eq_Hn}%
\end{align}
with%
\begin{align}
V_{0}  &  =M(\mathbf{k})-k_{A}^{2}(\frac{1+\eta^{2}}{2}M_{2}+\frac{M_{1}%
-M_{2}}{2}\sin^{2}\theta)\text{ },\nonumber\\
V_{1}  &  =k_{A}[M_{1}k_{z}\sin\theta-M_{2}k_{x}(\cos\theta\cos\phi-i\eta
\sin\phi)-M_{2}k_{y}(\cos\theta\sin\phi+i\eta\cos\phi)]\text{ },\nonumber\\
V_{-1}  &  =k_{A}[M_{1}k_{z}\sin\theta-M_{2}k_{x}(\cos\theta\cos\phi+i\eta
\sin\phi)-M_{2}k_{y}(\cos\theta\sin\phi-i\eta\cos\phi)]\text{ }. \label{eq_Vn}%
\end{align}
From the above equations, we can further derive%
\begin{equation}
\lbrack\mathcal{H}_{1,0}^{s},\mathcal{H}_{0,1}^{s}]=2\eta v_{0}M_{1}k_{A}%
^{2}\sin\theta\left(
\begin{array}
[c]{cc}%
0 & k_{z}e^{is\phi}\\
k_{z}e^{-is\phi} & 0
\end{array}
\right)  -2\eta v_{0}M_{2}k_{A}^{2}\cos\theta\left(
\begin{array}
[c]{cc}%
0 & k_{x}+isk_{y}\\
k_{x}-isk_{y} & 0
\end{array}
\right)  -s\eta v_{0}^{2}k_{A}^{2}\cos\theta\left(
\begin{array}
[c]{cc}%
1 & 0\\
0 & -1
\end{array}
\right)  . \label{eq_Anc}%
\end{equation}

Inserting Eqs. (\ref{eq_Hn}) and (\ref{eq_Anc}) into
\begin{equation}
\mathcal{H}_{\mathrm{eff}}^{s}(\mathbf{k})=\mathcal{H}_{0,0}^{s}%
+\frac{[\mathcal{H}_{1,0}^{s},\mathcal{H}_{0,1}^{s}]}{\omega},
\end{equation}
the effective Hamiltonian $\mathcal{H}_{\mathrm{eff}}(\mathbf{k}%
)=\mathcal{H}_{\mathrm{eff}}^{+}(\mathbf{k})\oplus\mathcal{H}_{\mathrm{eff}%
}^{-}(\mathbf{k})$ finally takes the form as given by Eq. (3) of the main text.

\begin{figure}[ptb]
\centering\includegraphics[width=0.8\linewidth]{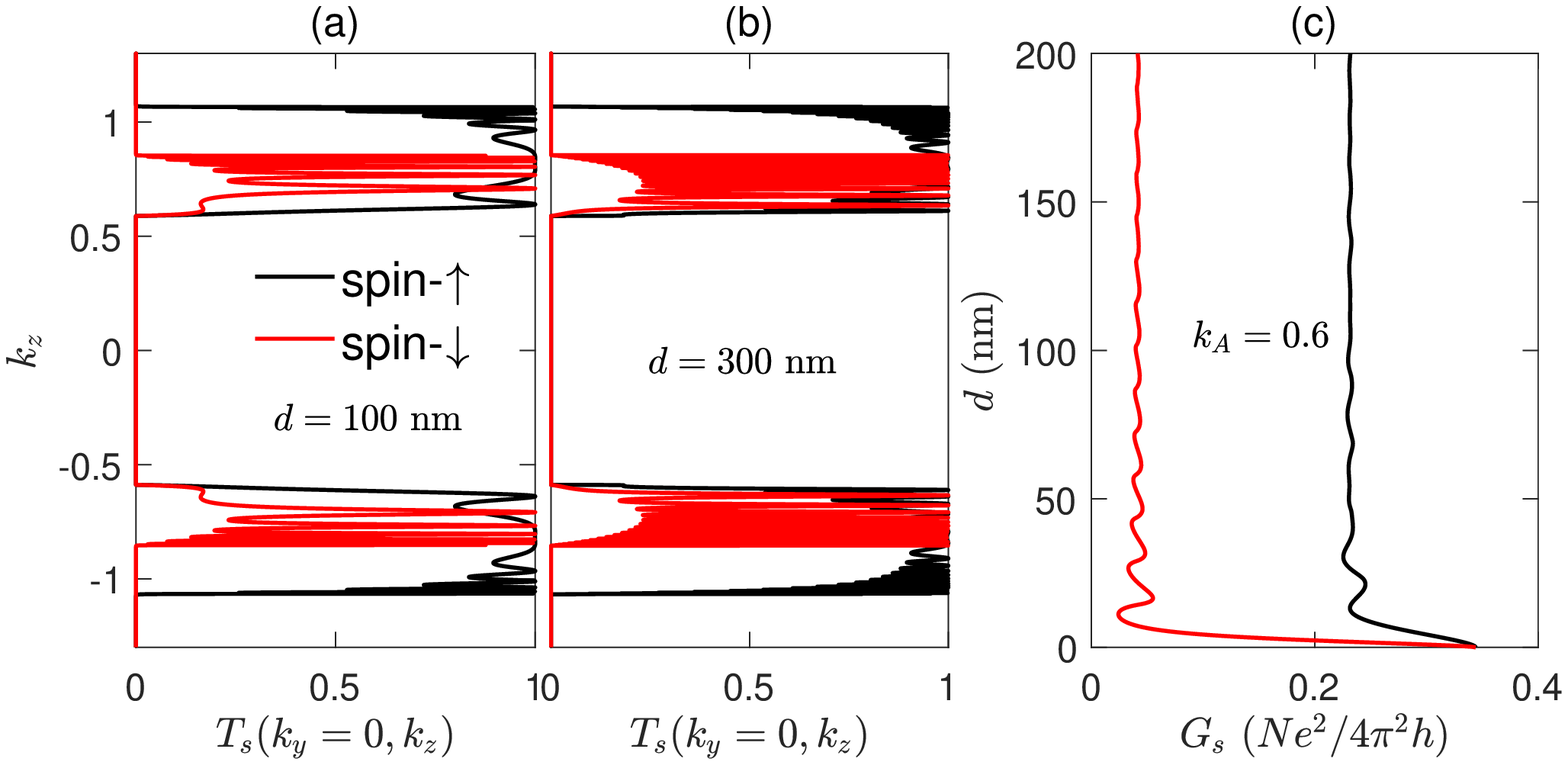} \caption{(a)-(b)
The spin-dependent transmission probability $T_{s}(k_{y}=0,k_{z})$ vs $k_{z}$
for $k_{A}=0.6$ \textrm{nm}$^{-1}$, with (a) $d=100$ \textrm{nm} and (b)
$d=300$ \textrm{nm}. (c) The spin-dependent conductances as functions of the
width $d$ of the junction. }%
\label{figS_TG}%
\end{figure}

\begin{center}
\textbf{Supplementary material B: Effect of $d$ on the transmission
probability and conductance }
\end{center}

Our numerical calculations show that, while the pattern of $T_{s}(k_{y}%
,k_{z})$ in the $k_{y}$-$k_{z}$ plane changes with $d$, the conductance is
insensitive to the width $d$ of the junction when $d$ is large enough. This
can be understood as follows. With the transmission coefficient Eq. (9) in the
main text, we can obtain the transmission probability as
\begin{equation}
T_{s}(k_{y},k_{z})=|t_{s}|^{2}=\frac{4\mathcal{F}_{1}^{2}\mathcal{F}_{2}^{2}%
}{4\mathcal{F}_{1}^{2}\mathcal{F}_{2}^{2}+\left[  (\mathcal{F}_{1}%
^{2}-\mathcal{F}_{2}^{2})^{2}+2\Delta_{y}^{2}(\mathcal{F}_{1}^{2}%
+\mathcal{F}_{2}^{2})+\Delta_{y}^{4}\right]  \sin^{2}(k_{2}d)}, \label{eq_Ts}%
\end{equation}
where $k_{2}$, being function of $k_{y}$ and $k_{z}$, i.e.,
\begin{equation}
k_{2}=\frac{1}{v_{s}}\sqrt{E^{2}-|\tilde{M}(\mathbf{k})-s\lambda|^{2}%
-v_{s}^{2}(k_{y}+s\frac{v_{1}}{v_{s}}k_{z}\sin\phi)^{2}},
\end{equation}
is spin dependent. As shown by Eq. (\ref{eq_Ts}), for $k_{2}=n\pi/d$,
$T_{s}(k_{y},k_{z})=1$ exhibits perfect transmission, which is attributed to
the well-known Klein tunneling effect
\cite{PhysRevLett.102.026807,Young:2009aa}.

From Eq. (\ref{eq_Ts}), one can see that the transmission probability
$T_{s}(k_{y},k_{z})$ depends on $d$ only through the trigonometric function
$\sin^{2}(k_{2}d)$. Obviously, the increase of $d$ would modify the
oscillation period of $\sin^{2}(k_{2}d)$ and thus affect $T_{s}(k_{y},k_{z})$
drastically, as shown in Fig. 3 of the main text. To see this more clearly, we
plot $T_{s}(k_{y}=0,k_{z})$ as functions of $k_{z}$ in Figs. \ref{figS_TG}(a)
and (b). For $d$ large enough, the increase of $d$ only increases the number
of the high-frequency oscillations in the overlap regions but does not change
the average value of the transmission probability. Therefore, after integral
over $k_{y}$ and $k_{z}$, both conductances $G_{\uparrow}$ and $G_{\downarrow
}$ almost remain a constant value. In order to verify this, we also plot the
dependence of $G_{s}$ on $d$ in Fig. \ref{figS_TG}(c). Even though $G_{s}$
oscillates for small $d$, the oscillations, for large $d$ ($>80nm$), become
negligible and so the conductances almost remain a constant value.

\begin{center}
\textbf{Supplementary material C: Discussions about the role of an applied
magnetic field/magnetic perturbation or a linearly polarized light}
\end{center}

For a linearly polarized light, the vector potential for the light can be
chosen as $\mathbf{A}=A_{0}\cos(\omega t)\mathbf{e}_{1}$, i.e., $\eta=0$. In
this case, both $\lambda$ and $v_{1,2}$ vanish in Eq. (3) of the main text, so
that the spin degeneracy always remains, seeing Eq. (4) in the main text.
Therefore, no WHM phase appears. If the system is subjected to an applied
magnetic field/magnetic perturbation, the Weyl fermions is described by the
Hamiltonian $H(\mathbf{k})+\sum_{i=x,y,z}m_{i}\sigma_{i}\tau_{0}$. The
magnetization along $x/y$ direction will mix two spins and no spin
polarization appears. For the magnetization along $z$ direction $\mathbf{m}%
=m_{z}\hat{e}_{z}$, the quasienergy spectrum of the Hamiltonian is
\begin{equation}
\epsilon_{s,\pm}(\mathbf{k})=\epsilon_{0}(\mathbf{k})+sm_{z}\pm\sqrt
{|M(\mathbf{k})|^{2}+v_{0}^{2}(k_{x}^{2}+k_{y}^{2})}.
\end{equation}
From here, one can see that $m_{z}$ only shifts the spectrums of
$\epsilon_{\uparrow,\pm}(\mathbf{k})$ and $\epsilon_{\downarrow,\pm
}(\mathbf{k})$ to opposite directions in energy, but cannot realize the
creation and annihilation of the Weyl node pairs. Therefore, an applied
magnetic field/magnetic perturbation or a linearly polarized light can not
achieve the interesting WHM phase.

\begin{center}
\textbf{Supplementary material D: Model Realization}
\end{center}

\textbf{1. Floquet theory in lattice model}

Considering the spatial dependence or spatial confinement of the light, we
first derive for the effective lattice model Hamiltonian. For the minimal
model in Eq. (1) of the main text, the mapping to a cuboid lattice is obtained
by the substitutions $k_{i}\rightarrow\frac{1}{a_{i}}\sin(k_{i}a_{i})$ and
$k_{i}^{2}\rightarrow\frac{2}{a_{i}^{2}}[1-\cos(k_{i}a_{i})]$. Performing the
Fourier transformation, we obtain the following static tight-binding
Hamiltonian
\begin{equation}
H_{\mathrm{stat}}=\sum_{i}c_{i}^{\dag}\epsilon_{0}c_{i}+\sum_{ij}c_{i}^{\dag
}t_{ij}c_{j}+h.c.,
\end{equation}
where $\epsilon_{0}=\left(  M_{0}-\frac{2M_{1}}{a_{\bot}^{2}}-\frac{4M_{2}%
}{a_{\Vert}^{2}}\right)  \sigma_{z}$ and
\begin{equation}
t_{ij}=\left(
\begin{array}
[c]{cc}%
\frac{M_{2}}{a_{\Vert}^{2}} & -is\frac{v_{0}}{2a_{\Vert}}\\
-is\frac{v_{0}}{2a_{\Vert}} & -\frac{M_{2}}{a_{\Vert}^{2}}%
\end{array}
\right)  \delta_{i,j-\hat{x}a_{\Vert}}+\left(
\begin{array}
[c]{cc}%
\frac{M_{2}}{a_{\Vert}^{2}} & \frac{v_{0}}{2a_{\Vert}}\\
-\frac{v_{0}}{2a_{\Vert}} & -\frac{M_{2}}{a_{\Vert}^{2}}%
\end{array}
\right)  \delta_{i,j-\hat{y}a_{\Vert}}+\left(
\begin{array}
[c]{cc}%
\frac{M_{1}}{a_{\bot}^{2}} & 0\\
0 & -\frac{M_{1}}{a_{\bot}^{2}}%
\end{array}
\right)  \delta_{i,j-\hat{z}a_{\bot}}%
\end{equation}
with $a_{x,y}=a_{\Vert}$ and $a_{z}=a_{\bot}$ being lattice constants in the
$x$-$y$ plane and along the $z$ direction, respectively. The trivial term
$\epsilon_{0}(\mathbf{k})$ is omitted for the sake of brevity. For the central
region, where the fermions are driven by the CPL, the time-dependent
tight-binding Hamiltonian $H(t)$ is obtained from $H_{\mathrm{stat}}$ by
substituting $t_{ij}\rightarrow e^{iA_{ij}}t_{ij}$, where
\begin{equation}
A_{ij}=\frac{e}{\hbar}\int_{\mathbf{r}_{j}}^{\mathbf{r}_{i}}\mathbf{A}%
(\mathbf{r},t)\cdot d\mathbf{l}%
\end{equation}
is phase change of the hopping integral due to the CPL. Subsequently, we can
derive%
\begin{align}
H^{s}(t)  &  =\sum_{i}c_{i}^{\dag}\epsilon_{0}c_{i}+\sum_{i}\left[
e^{i\zeta_{x}\cos\left(  \omega t+\eta\alpha_{x}\right)  }c_{i}^{\dag}\left(
\begin{array}
[c]{cc}%
\frac{M_{2}}{a_{\Vert}^{2}} & -is\frac{v_{0}}{2a_{\Vert}}\\
-is\frac{v_{0}}{2a_{\Vert}} & -\frac{M_{2}}{a_{\Vert}^{2}}%
\end{array}
\right)  c_{i+\hat{x}a_{\Vert}}+h.c.\right] \nonumber\\
&  +\sum_{i}\left[  e^{i\zeta_{y}\cos\left(  \omega t-\eta\alpha_{y}\right)
}c_{i}^{\dag}\left(
\begin{array}
[c]{cc}%
\frac{M_{2}}{a_{\Vert}^{2}} & \frac{v_{0}}{2a_{\Vert}}\\
-\frac{v_{0}}{2a_{\Vert}} & -\frac{M_{2}}{a_{\Vert}^{2}}%
\end{array}
\right)  c_{i+\hat{y}a_{\Vert}}+h.c.\right] \nonumber\\
&  +\sum_{i}\left[  e^{-i\zeta_{z}\cos\omega t}c_{i}^{\dag}\left(
\begin{array}
[c]{cc}%
\frac{M_{1}}{a_{\bot}^{2}} & 0\\
0 & -\frac{M_{1}}{a_{\bot}^{2}}%
\end{array}
\right)  c_{i+\hat{z}a_{\bot}}+h.c.\right]  ,
\end{align}
where
\begin{align}
\zeta_{x}  &  =k_{A}a_{\Vert}\sqrt{\cos^{2}\theta\cos^{2}\phi+\sin^{2}\phi
}\nonumber\\
\zeta_{y}  &  =k_{A}a_{\Vert}\sqrt{\cos^{2}\theta\sin^{2}\phi+\cos^{2}\phi
}\nonumber\\
\zeta_{z}  &  =k_{A}a_{\bot}\sin\theta
\end{align}
and%
\begin{align}
\sin\alpha_{x}  &  =\frac{\sin\phi}{\sqrt{\cos^{2}\theta\cos^{2}\phi+\sin
^{2}\phi}}\nonumber\\
\sin\alpha_{y}  &  =\frac{\cos\phi}{\sqrt{\cos^{2}\theta\sin^{2}\phi+\cos
^{2}\phi}}.
\end{align}
Now, the spacial dependence of the vector potential can be taken into account,
i.e.,
\begin{equation}
k_{A}=\frac{eA_{0}}{\hbar}\cos\left(  \frac{2\pi}{\lambda_{c}}\mathbf{n}%
\cdot\mathbf{r}\right)  =\frac{eA_{0}}{\hbar}\cos\left(  2\pi\frac{x\sin
\theta\cos\phi+y\sin\theta\sin\phi+z\cos\theta}{\lambda_{c}}\right)
\label{eq_kA}%
\end{equation}
with $\lambda_{c}$ the wavelength of the CPL. By using the relation
\begin{equation}
e^{ix\cos\xi}=\sum_{n=-\infty}^{\infty}(i)^{n}J_{n}(x)e^{in\xi}%
\end{equation}
and performing the Fourier transformation on the basis and time, we can
further derive the photodressed lattice Hamiltonian
\begin{align}
\mathcal{H}_{00}^{s}  &  =M^{\prime}(\mathbf{k})\sigma_{z}+s\frac{J_{0,x}%
v_{0}}{a_{\Vert}}\sin(k_{x}a_{\Vert})\sigma_{x}-\frac{J_{0,y}v_{0}}{a_{\Vert}%
}\sin(k_{y}a_{\Vert})\sigma_{y}\nonumber\\
\mathcal{H}_{1,0}^{s}  &  =M_{10}(\mathbf{k})\sigma_{z}+se^{-i\eta\alpha_{x}%
}\frac{J_{1,x}v_{0}}{a_{\Vert}}\cos(k_{x}a_{\Vert})\sigma_{x}-e^{i\eta
\alpha_{y}}\frac{J_{1,y}v_{0}}{a_{\Vert}}\cos(k_{y}a_{\Vert})\sigma
_{y}\nonumber\\
\mathcal{H}_{0,1}^{s}  &  =M_{01}(\mathbf{k})\sigma_{z}+se^{i\eta\alpha_{x}%
}\frac{J_{1,x}v_{0}}{a_{\Vert}}\cos(k_{x}a_{\Vert})\sigma_{x}-e^{-i\eta
\alpha_{y}}\frac{J_{1,y}v_{0}}{a_{\Vert}}\cos(k_{y}a_{\Vert})\sigma_{y},
\end{align}
where%
\begin{align}
M^{\prime}(\mathbf{k})  &  =M_{0}-\sum_{i}\frac{2M_{i}}{a_{i}^{2}}+\sum
_{i}\frac{2J_{0,i}M_{i}}{a_{i}^{2}}\cos(k_{i}a_{i})\nonumber\\
M_{10}(\mathbf{k})  &  =-e^{-i\eta\alpha_{x}}\frac{2J_{1,x}M_{2}}{a_{\Vert
}^{2}}\sin(k_{x}a_{\Vert})-e^{i\eta\alpha_{y}}\frac{2J_{1,y}M_{2}}{a_{\Vert
}^{2}}\sin(k_{y}a_{\Vert})+\frac{2J_{1,z}M_{1}}{a_{\bot}^{2}}\sin(k_{z}%
a_{\bot})\nonumber\\
M_{01}(\mathbf{k})  &  =-e^{i\eta\alpha_{x}}\frac{2J_{1,x}M_{2}}{a_{\Vert}%
^{2}}\sin(k_{x}a_{\Vert})-e^{-i\eta\alpha_{y}}\frac{2J_{1,y}M_{2}}{a_{\Vert
}^{2}}\sin(k_{y}a_{\Vert})+\frac{2J_{1,z}M_{1}}{a_{\bot}^{2}}\sin(k_{z}%
a_{\bot})
\end{align}
with $M_{z}=M_{1}$, $M_{x,y}=M_{2}$, $J_{n,i}=J_{n}(\zeta_{i})$ and $J_{n}(x)$
the $n$-$\mathrm{th}$ order Bessel function. In the off-resonance regime, the
Bessel functions $J_{n}(x)$ with $n>1$ are negligible. Therefore, the
effective lattice Hamiltonian for the driving region can be approximated as
\begin{align}
H_{\mathrm{eff}}^{s}  &  =\mathcal{H}_{0,0}^{s}+\frac{[\mathcal{H}_{1,0}%
^{s},\mathcal{H}_{0,1}^{s}]}{\omega}\nonumber\\
&  =\left[  M^{\prime}(\mathbf{k})-s\eta\frac{2v_{0}^{2}}{a_{\Vert}^{2}}%
\frac{J_{1,x}J_{1,y}}{\omega}\sin(\alpha_{x}+\alpha_{y})\cos(k_{x}a_{\Vert
})\cos(k_{y}a_{\Vert})\right]  \sigma_{z}\nonumber\\
&  +\left[  s\frac{J_{0,x}v_{0}}{a_{\Vert}}-\eta\frac{4M_{2}v_{0}}{a_{\Vert
}^{3}}\frac{J_{1,x}J_{1,y}}{\omega}\sin(\alpha_{x}+\alpha_{y})\cos
(k_{y}a_{\Vert})\right]  \sin(k_{x}a_{\Vert})\sigma_{x}\nonumber\\
&  -\left[  \frac{J_{0,y}v_{0}}{a_{\Vert}}-s\eta\frac{4M_{2}v_{0}}{a_{\Vert
}^{3}}\frac{J_{1,x}J_{1,y}}{\omega}\sin(\alpha_{x}+\alpha_{y})\cos
(k_{x}a_{\Vert})\right]  \sin(k_{y}a_{\Vert})\sigma_{y}\nonumber\\
&  +\eta\frac{4M_{1}v_{0}}{a_{\Vert}a_{\bot}^{2}}\left[  \frac{J_{1,y}J_{1,z}%
}{\omega}\sin\alpha_{y}\cos(k_{y}a_{\Vert})\sigma_{x}-s\frac{J_{1,x}J_{1,z}%
}{\omega}\sin\alpha_{x}\cos(k_{x}a_{\Vert})\sigma_{y}\right]  \sin
(k_{z}a_{\bot}). \label{eq_sHeff}%
\end{align}
Subsequently, we obtain the following tight-binding Hamiltonian for the
driving region%
\begin{equation}
H_{\mathrm{D}}^{s}=\sum_{i}c_{i}^{\dag}\epsilon_{0}c_{i}+\sum_{ij}c_{i}^{\dag
}\left(  t_{1,ij}+t_{2,ij}\right)  c_{j}+h.c., \label{eq_sHdr}%
\end{equation}
where%
\begin{equation}
t_{1,ij}=J_{0,x}\left(
\begin{array}
[c]{cc}%
\frac{M_{2}}{a_{\Vert}^{2}} & -is\frac{v_{0}}{2a_{\Vert}}\\
-is\frac{v_{0}}{2a_{\Vert}} & -\frac{M_{2}}{a_{\Vert}^{2}}%
\end{array}
\right)  \delta_{i,j-\hat{x}a_{\Vert}}+J_{0,y}\left(
\begin{array}
[c]{cc}%
\frac{M_{2}}{a_{\Vert}^{2}} & \frac{v_{0}}{2a_{\Vert}}\\
-\frac{v_{0}}{2a_{\Vert}} & -\frac{M_{2}}{a_{\Vert}^{2}}%
\end{array}
\right)  \delta_{i,j-\hat{y}a_{\Vert}}+J_{0,z}\left(
\begin{array}
[c]{cc}%
\frac{M_{1}}{a_{\bot}^{2}} & 0\\
0 & -\frac{M_{1}}{a_{\bot}^{2}}%
\end{array}
\right)  \delta_{i,j-\hat{z}a_{\bot}}%
\end{equation}
and%
\begin{align}
t_{2,ij}  &  =\left(  iv_{2}\sigma_{x}-isv_{2}\sigma_{y}-s\lambda\sigma
_{z}\right)  \delta_{i,j-\hat{x}a_{\Vert}-\hat{y}a_{\Vert}}+\left(
iv_{2}\sigma_{x}+isv_{2}\sigma_{y}-s\lambda\sigma_{z}\right)  \delta
_{i,j-\hat{x}a_{\Vert}+\hat{y}a_{\Vert}}\nonumber\\
&  +isv_{x}\sigma_{y}\delta_{i,j-\hat{x}a_{\Vert}-\hat{z}a_{\bot}}%
-isv_{x}\sigma_{y}\delta_{i,j-\hat{x}a_{\Vert}+\hat{z}a_{\bot}}-iv_{y}%
\sigma_{x}\delta_{i,j-\hat{y}a_{\Vert}-\hat{z}a_{\bot}}+iv_{y}\sigma_{x}%
\delta_{i,j-\hat{y}a_{\Vert}+\hat{z}a_{\bot}} \label{eq_t2ij}%
\end{align}
with
\begin{align}
\lambda &  =\frac{\eta v_{0}^{2}}{2a_{\Vert}^{2}}\frac{J_{1,x}J_{1,y}}{\omega
}\sin(\alpha_{x}+\alpha_{y})\nonumber\\
v_{2}  &  =\frac{\eta M_{2}v_{0}}{a_{\Vert}^{3}}\frac{J_{1,x}J_{1,y}}{\omega
}\sin(\alpha_{x}+\alpha_{y})\nonumber\\
v_{x,y}  &  =\frac{\eta M_{1}v_{0}}{a_{\Vert}a_{\bot}^{2}}\frac{J_{1,x,y}%
J_{1,z}}{\omega}\sin\alpha_{x,y}.
\end{align}
As shown by Eqs. (\ref{eq_sHdr}) and (\ref{eq_t2ij}), the off-resonant CPL, on
one hand, modifies the nearest hopping integral and, on the other hand,
introduces second nearest hopping to the lattice Hamiltonian.

\textbf{2. Effect of spatially periodic dependence of the vector potential
along propagation direction of the light}
\begin{figure}[ptb]
\centering\includegraphics[width=0.8\linewidth]{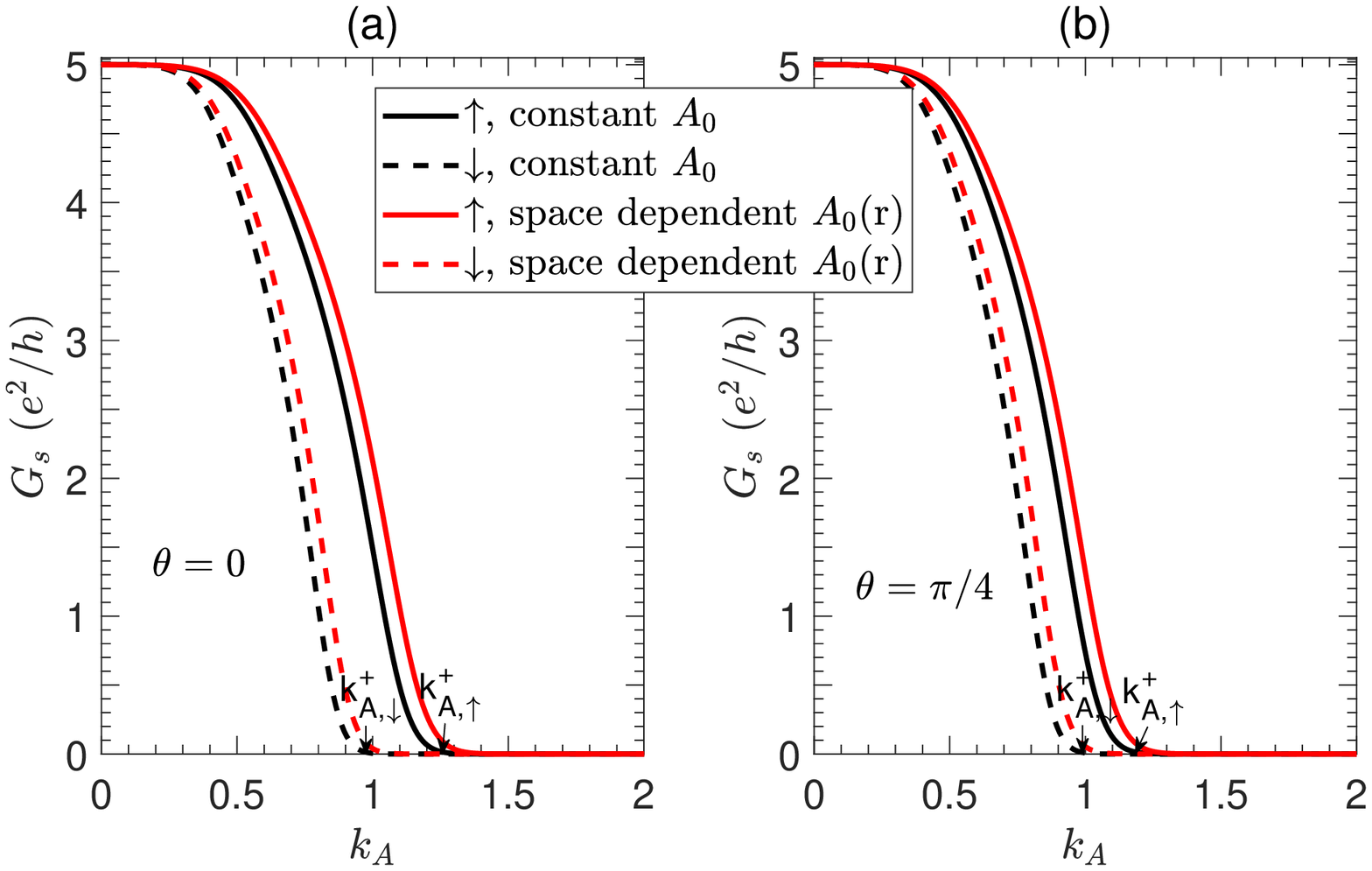} \caption{ Effect
of spatially periodic dependence of the vector potential on the spin-dependent
conductances $G_{s}$ with $\phi=\pi/2$, and (a) $\theta=0$ and (b) $\theta
=\pi/4$. The lattice size of the junction is chosen as $20a_{\Vert}%
\times20a_{\Vert}\times20a_{\bot}$ with $a_{\Vert}=a_{\bot}=a$, and
$\lambda_{c}=90a$ for numerical calculations using KWANT\cite{Groth2014}.}%
\label{figS_Gs}%
\end{figure}

As we know, when light propagates along an incident direction, the vector
potential is spatially periodic, which takes the form $\mathbf{A}%
=A_{0}(\mathbf{r})[\cos(\omega t)\mathbf{e}_{1}-\eta\sin(\omega t)\mathbf{e}%
_{2}]$, where
\begin{equation}
A_{0}(\mathbf{r})=A_{0}\cos\left(  \frac{2\pi}{\lambda_{c}}\mathbf{n}%
\cdot\mathbf{r}\right)  =A_{0}\cos\left(  2\pi\frac{x\sin\theta\cos\phi
+y\sin\theta\sin\phi+z\cos\theta}{\lambda_{c}}\right)
\end{equation}
with $\mathbf{e}_{1,2}$ being two unit vectors perpendicular to the incident
direction $\mathbf{n}$. When taking into account the spacial inhomogeneity of
the vector potential, the photodressed lattice Hamiltonian is given by Eq.
(A23). Obviously, the problems are unable to be solved analytically. In this
discussion, we employ the widely adopted transport package
KWANT\cite{Groth2014}, which is convenient to deal with this problem
numerically. For specifical calculations, the size of the sample is taken to
be $20a_{\Vert}\times20a_{\Vert}\times20a_{\bot}$ with $a_{\Vert}=a_{\bot}=a$,
and the calculated results for spin-resolved conductances $G_{s}$ are plotted
in Fig. \ref{figS_Gs}. Compared the results with (red lines) and without
(black lines) the spacial inhomogeneity, one can find that the spacial
inhomogeneity due to propagation of the light only shifts the critical values
of $k_{A}$, but does not change the spin filter effect qualitatively, whether
for vertical ($\theta=0$) or oblique ($\theta=\pi/4$) incidence. This
influence can be greatly reduced if keeping $x,y,z\ll\lambda_{c}$.

Experimentally, the sample can be fabricated as a microribbon as illustrated
in Fig. 1(b) of the main text, with several hundreds of nanometer in width
($y$ direction) and thickness ($z$ direction), and several micrometers in
length ($x$ direction). Since the thickness and width of the sample are much
smaller than the wavelength of the light, the spatial inhomogeneity of the
vector potential along $y$ and $z$ directions can be ignored safely. At the
same time, if choosing the incident direction $\mathbf{n}$ with $\phi
=\frac{\pi}{2}$ (i.e., tilting in the $y$-$z$ plane), or confining the
irradiation region $d$ in several hundreds of nanometer, the spatially
periodic dependence of the vector potential along the junction ($x$ direction)
also can be negligible. On the other hand, although the size of several
hundreds of nanometer is far smaller than the wavelength of the light, it is
greatly larger than the Bloch wavelength of the electrons. Thus, the numerical
results calculated with the lattice model, i.e., above Fig. 2, are
qualitatively consistent with the analytical results calculated with the
continuous model, i.e., Fig. 4 in the main text. In experiments, one should
irradiate the light along $z$ ($\theta=0$), where the spin filter effect is
strongest and the influence of the spatial inhomogeneity is weakest.

\begin{figure}[ptb]
\centering\includegraphics[width=0.5\linewidth]{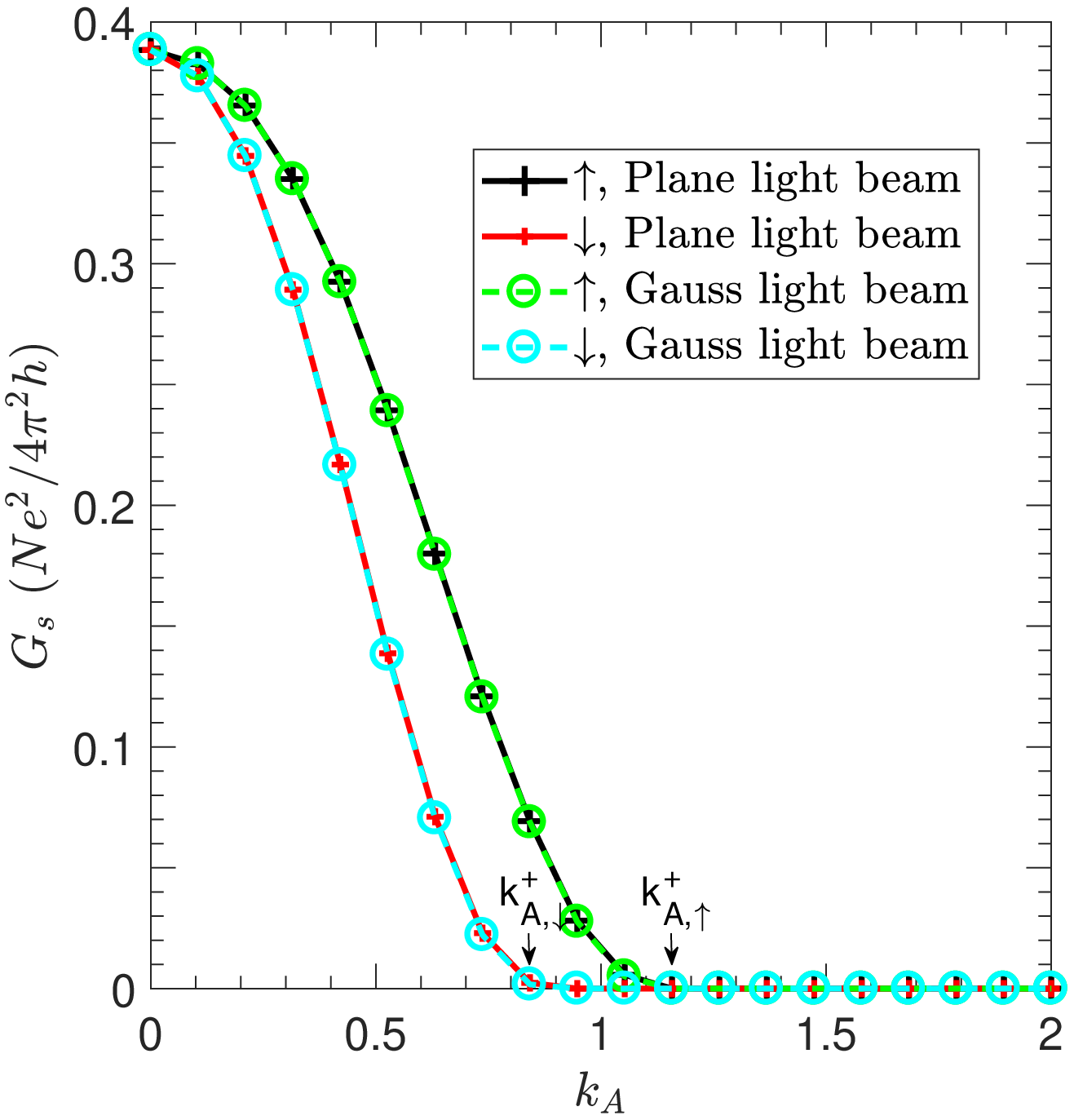} \caption{
Conductances with the incident plane light beam and the Gauss light beam
irradiating from vertical direction $\theta=0$. The Gauss distribution of the
light intensity takes the form of $A_{0}\rightarrow A_{0}(\mathbf{r}%
)=\frac{A_{0}}{D}exp[\frac{-(x^{2}+y^{2})}{D^{2}}]$, where we set the spot
radius $D=d/2$ locating at the boundaries between the irradiation region and
reservoirs. The length of the irradiation region is set as $d=500a$ for
numerical calculations using KWANT\cite{Groth2014}.}%
\label{figS_Tk}%
\end{figure}

\textbf{3. Realization of the model with light confinement}

In the model of Fig. 1(b) in the main text, we assume that the light intensity
is uniform inside the junction and drops to zero outside. Experimentally, to
confine the light in the width of $d$ and to reach the most prominent spin
filter effect, one can irradiate a beam of gaussian-profile light along $z$
direction ($\theta=0$). In this situation, the vector potential $A_{0}$ in Eq.
(2) of the main text is modified to $A_{0}\rightarrow A_{0}(\mathbf{r}%
)=\frac{A_{0}}{D}exp[\frac{-(x^{2}+y^{2})}{D^{2}}]$, where we set the spot
radius $D=d/2$ locating at the boundaries between the irradiation region and reservoirs.

Including the spatial dependence of the light intensity, we perform numerical
calculations with the lattice model. For simplicity, we neglect spatially
periodic dependence of the vector potential along propagation direction $z$.
Taking $k_{y}$ and $k_{z}$ as parameters, we map the continuous Hamiltonian to
a one-dimensional chain along direction $x$ and then perform numerical
calculations using KWANT. The numerical results are presented in Fig.
\ref{figS_Tk}. As seen from Fig. \ref{figS_Tk}, the conductances calculated
with the gaussian-profile light beam (with $A_{0}(\mathbf{r})$) exhibit almost
the same spin filter effect as that calculated with the plane light beam
(constant $A_{0}$). This, in fact, can be expected, since the Gauss
distribution of the light intensity is mainly to renormalize the effective
width of the junction, while as discussed in supplementary material B, the
conductances are independent on the width of the junction for the chosen light
beam. These numerical results are also consistent with the analytical
calculations in the main text.

\textbf{4. Realization of the model without light confinement}

In fact, we can loose the restriction of focusing the light on the central
region for the model in Fig. 1(b) of the main text. Alternatively, in this
setup, one can replace two Weyl-semimetal reservoirs with two normal-metal
electrodes, i.e., fabricating a normal metal/Weyl semimetal/normal metal
junction. The light can irradiate on the whole system. In order to reach the
strongest spin filter effect, one should irradiate the light along $z$
($\theta=0$). Due to trivial response of the normal-metal electrodes to the
irradiation, the transport is determined by the WHM phase.

For this normal metal/Weyl semimetal/normal metal junction, the effective
tight-binding Hamiltonian for the middle region is given by Eq. (\ref{eq_sHdr}%
) and the continuous Hamiltonian for the normal metal leads is $H_{lead}%
=\frac{\hbar^{2}k^{2}}{2m}-\mu$, with $m$ the mass and $\mu$ the chemical
potential. Following the procedure in piont 1 of supplementary material D, we
first obtain the static tight-binding Hamiltonian for the normal metal leads
as%
\begin{equation}
H_{lat}=\sum_{i}\left(  -\mu+6t\right)  c_{i}^{\dag}c_{i}-\left(
t\sum_{\left\langle ij\right\rangle }c_{i}^{\dag}c_{j}+h.c.\right)
\end{equation}
with $t=\frac{\hbar^{2}}{2ma^{2}}$ the nearest neighbouring hopping integral
and $a$ the lattice constant. In the presence of the CPL, the effective
lattice Hamiltonian for the normal metal leads can be derived as
\begin{equation}
H_{lat}^{\mathrm{eff}}=\sum_{i}\left(  -\mu+6t\right)  c_{i}^{\dag}c_{i}%
-t\sum_{i}\left(  J_{0,x}c_{i}^{\dag}c_{i+\hat{x}a}+J_{0,y}c_{i}^{\dag
}c_{i+\hat{y}a}+J_{0,z}c_{i}^{\dag}c_{i+\hat{z}a}+h.c.\right)  ,
\label{eq_Hlat}%
\end{equation}
from which we can see that the response of the normal metal leads to the CPL
is quite trivial, namely, although the nearest neighboring hopping integral
$t$ is renormalized, no additional hopping terms are introduced.

\begin{figure}[ptb]
\centering\includegraphics[width=0.4\linewidth]{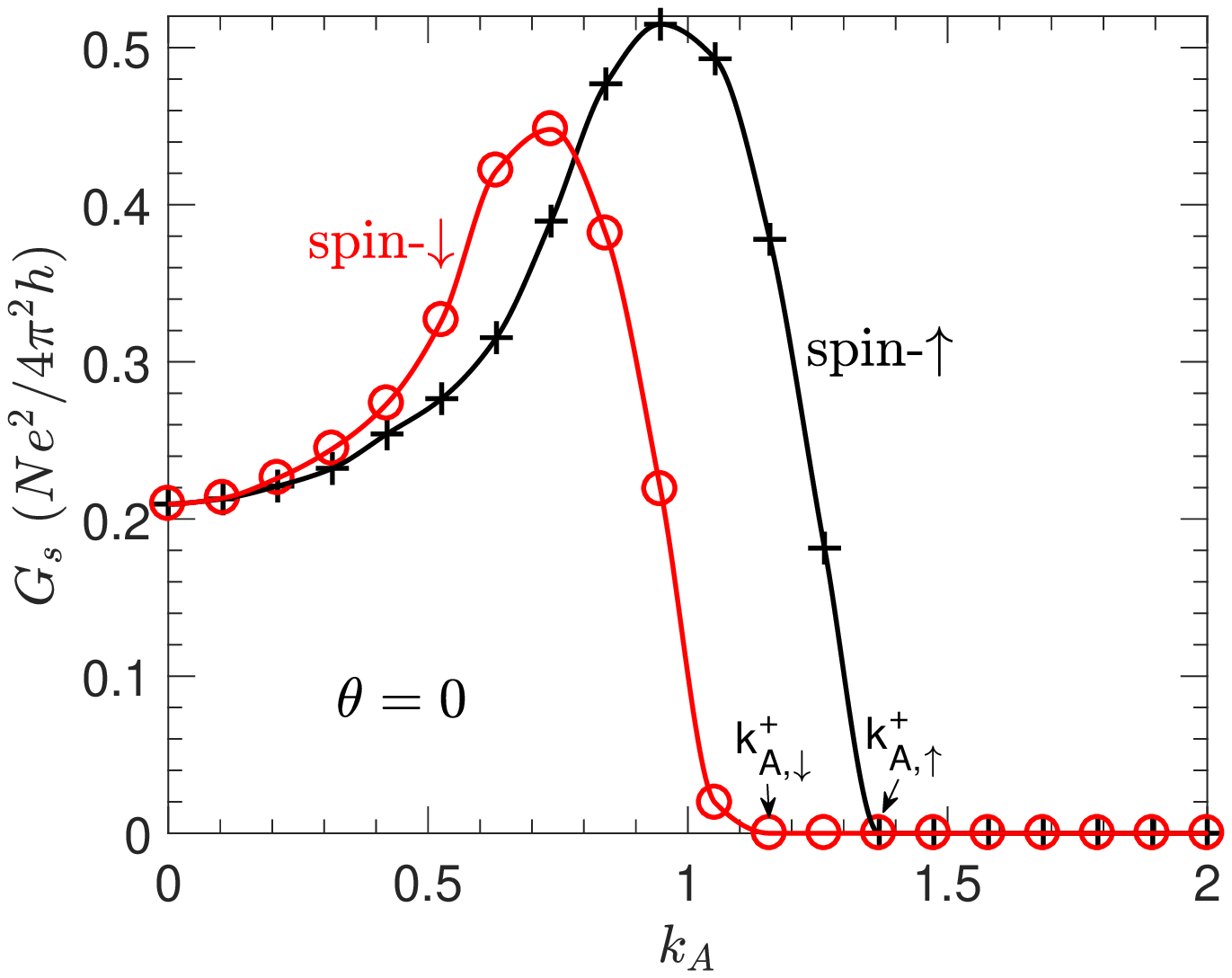}\caption{Spin-dependent
conductances in a normal metal/Weyl semimetal/normal metal junction as
functions of the CPL intensity $k_{A}$, with $\theta=0$, $t=0.1$ $\mathrm{eV}%
$, and $\mu=0.1$ $\mathrm{eV}$. Other parameters chosen the same as Fig.
\ref{figS_Tk}.}%
\end{figure}

Performing the Fourier transform on Eqs. (\ref{eq_sHdr}) along the $y$ and $z$
directions and ignoring the spatial dependence of CPL, we can approximate%
\begin{equation}
H_{\mathrm{D}}^{s}=\sum_{i\in x_{m}/a}\left[  c_{i}^{\dag}\epsilon_{0}%
^{\prime}c_{i}+\left(  c_{i}^{\dag}t_{m}c_{i+1}+h.c.\right)  \right]  ,
\end{equation}
in the vicinity of the gap closing points for the middle region $x_{m}%
\in\lbrack0,d]$, with
\begin{align}
\epsilon_{0}^{\prime}  &  =4v_{y}\sin(k_{z}a)\sigma_{x}-\left[  \left(
\frac{J_{0,y}v_{0}}{a}-4sv_{2}\right)  \sin(k_{y}a)+4sv_{x}\sin(k_{z}%
a)\right]  \sigma_{y}\nonumber\\
&  +\left[  M_{0}-\frac{2M_{1}}{a^{2}}-\frac{4M_{2}}{a^{2}}+\frac
{2J_{0,y}M_{2}}{a^{2}}\cos(k_{y}a)+\frac{2J_{0,z}M_{2}}{a^{2}}\cos
(k_{z}a)-4s\lambda\right]  \sigma_{z},
\end{align}
and%
\begin{equation}
t_{m}=\frac{J_{0,x}M_{2}}{a^{2}}\sigma_{z}-\frac{i}{2}\left(  s\frac
{J_{0,x}v_{0}}{a}-4v_{2}\right)  \sigma_{x}.
\end{equation}

Similarly, for the normal metal leads with $x_{L}\in(-\infty,-a]\cup\lbrack
d+a,+\infty)$, we can derive the one-dimensional lattice Hamiltonian as%
\begin{equation}
H_{lat}^{\mathrm{eff}}=\sum_{i\in x_{L}/a}\left[  -\mu+6t-2tJ_{0,y}\cos
(k_{y}a)-2tJ_{0,z}\cos(k_{z}a)\right]  c_{i}^{\dag}c_{i}-t\sum_{i\in x_{L}%
/a}\left(  J_{0,x}c_{i}^{\dag}c_{i+1}+h.c.\right)  . \label{eq_1Hlat}%
\end{equation}
The coupling Hamiltonian between the middle region and two normal-metal leads
can be obtained via the last term of Eq. (\ref{eq_1Hlat}) by fixing $i=-1$ and
$d/a$.

We carry out the numerical calculations using KWANT and the numerical results
are presented in Fig. 4. As expected, figure 4 shows the prominent spin filter
effect. Compared with the Weyl-semimetal electrodes, the difference here is
that there is a drop below the spin filter region. It is easy to be
understood. Without the CPL, the Fermi surface of Weyl-semimetal electrodes is
completely overlapped with one in the middle region, and so the conductance is
largest. In contrast, the Fermi surface of the normal-metal electrodes is
located in the middle of two Weyl nodes, in which the overlapping area between
the middle region and the electrodes is small. With increase of the
illumination, the reduced distance between two Weyl nodes makes the
overlapping area first increase and then decrease due to open of the energy
gap band, which leads to nonmonotonical behavior in Fig. 4. In this process,
the effect of the CPL on the normal metal mainly renormalizes its chemical
potential, namely, changing the rate of descent below the spin-filter region
but never changes $k_{A,\sigma}^{+}$ because it is determined by phase
boundary of the WSM phase. In fact, the spin filter effect as a spin-related
transport, originating from the spin-polarized band in WHM phase, is quite
insensitive to the nonmagnetic electrodes and junction details.

\begin{center}
\textbf{Supplementary material E: Kwant scripts related to the numerical simulations}
\end{center}
In this section, we would attach the Kwant scripts related to the numerical simulations in supplemental material D as follows.

\textbf{1. Effect of spatially periodic dependence of the vector potential}

\begin{verbatim}
#!/usr/bin/env python3
# -*- coding: utf-8 -*-
"""
@author: Ming-Xun Deng
"""
import kwant

from matplotlib import pyplot

from scipy import special

import numpy as np

import tinyarray

# define Pauli-matrices for convenience
sigma_0 = tinyarray.array([[1, 0], [0, 1]])

sigma_x = tinyarray.array([[0, 1], [1, 0]])

sigma_y = tinyarray.array([[0, -1j], [1j, 0]])

sigma_z = tinyarray.array([[1, 0], [0, -1]])

# system parameters

a = 1

yita = 1

theta = 0

phi = np.pi / 2

omega = 1

Ef = 0.01

v0 = 0.24598

M0 = - 0.08686

M1 = - 0.106424

M2 = - 0.103610

L_c = 100  #  light wavelength

Gauss = 0  # 0 plane beam and > 0 for Gaussian light beam

L = 20

W = 20

H = 20

rho_x = np.sqrt( np.cos(theta)**2 * np.cos(phi)**2 + np.sin(phi)**2 )

rho_y = np.sqrt( np.cos(theta)**2 * np.sin(phi)**2 + np.cos(phi)**2 )

rho_z = np.sin(theta)

sin_ax = np.sin(phi) / rho_x

sin_ay = np.cos(phi) / rho_y

sin_xy = ( np.sin(phi) * np.cos(theta) * np.sin(phi) \
        + np.cos(theta) * np.cos(phi) * np.cos(phi) )/(rho_x * rho_y )

def make_system( s = 1, A0 = 0 ):
    # Start with an empty tight-binding system and a single square lattice.
    # `a` is the lattice constant (by default set to 1 for simplicity).

    lat = kwant.lattice.cubic( a )

    sys = kwant.Builder( )

    def hopping( h_type ):

        def hop_phi( sitei, sitej ):

            ( xi, yi, zi ) = sitei.pos

            ( xj, yj, zj ) = sitej.pos

            x = ( xi + xj ) / 2

            y = ( yi + yj ) / 2

            z = ( zi + zj ) / 2

            r_n = ( x - L / 2 )*np.sin(theta)*np.cos(phi) \
                 + ( y - W / 2 )*np.sin(theta)*np.sin(phi) + ( z- H /2 ) * np.cos(theta)

            R_G = ( x - L / 2 ) ** 2 + ( y - W / 2 ) ** 2 + ( z- H /2 ) **2 - r_n ** 2

            if np.abs(Gauss) > 0:
                pA =  np.exp( - R_G / Gauss ** 2  )
            else:
                pA = 1

            if np.abs( L_c ) > 0:
                kA = pA * A0 * np.cos( 2*np.pi * r_n / L_c  )
            else:
                kA = pA * A0

            J_0x = special.jv(0, kA * rho_x )

            J_0y = special.jv(0, kA * rho_y )

            J_0z = special.jv(0, kA * rho_z )

            J_1x = special.jv(1, kA * rho_x )

            J_1y = special.jv(1, kA * rho_y )

            J_1z = special.jv(1, kA * rho_z )

            v2 = yita * M2 * v0 * J_1x * J_1y * sin_xy / omega

            vx = yita * M1 * v0 * J_1x * J_1z * sin_ax / omega

            vy = yita * M1 * v0 * J_1y * J_1z * sin_ay / omega
            lam = yita * v0 **2 * J_1x * J_1y * sin_xy / ( 2 * omega )

            if h_type == 1 :
                return J_0x * ( M2 * sigma_z - s * 1j * v0 / 2 * sigma_x )
            elif h_type == 2 :
                return J_0y * ( M2 * sigma_z + 1j * v0 / 2 * sigma_y )
            elif h_type == 3:
                return J_0z * M1 * sigma_z
            elif h_type == 4:
                return 1j * v2 * sigma_x - 1j * s * v2 * sigma_y - s * lam * sigma_z
            elif h_type == 5:
                return 1j * v2 * sigma_x + 1j * s * v2 * sigma_y - s * lam * sigma_z
            elif h_type == 6:
                return 1j * s * vx * sigma_y
            elif h_type == 7:
                return - 1j * s * vx * sigma_y
            elif h_type == 8:
                return - 1j * vy * sigma_x
            else:
                return 1j * vy * sigma_x

        hopping_coeff = hop_phi

        return hopping_coeff

    #### Define the scattering region. ####
    sys[( lat(i, j, k) for i in range(L) for j in range(W) for k in range(H) )] = \
        ( M0 - 2 * M1 - 4 * M2 ) * sigma_z

    # hoppings in x-direction
    sys[kwant.builder.HoppingKind((1, 0, 0), lat, lat)] = hopping( 1 )
    # hoppings in y-directions
    sys[kwant.builder.HoppingKind((0, 1, 0), lat, lat)] = hopping( 2 )
    # hoppings in z-directions
    sys[kwant.builder.HoppingKind((0, 0, 1), lat, lat)] = hopping( 3 )

    ## second nearest hopping
    sys[kwant.builder.HoppingKind((1,  1,  0), lat, lat)] = hopping( 4 )
    sys[kwant.builder.HoppingKind((1, -1,  0), lat, lat)] = hopping( 5 )
    sys[kwant.builder.HoppingKind((1,  0,  1), lat, lat)] = hopping( 6 )
    sys[kwant.builder.HoppingKind((1,  0, -1), lat, lat)] = hopping( 7 )
    sys[kwant.builder.HoppingKind((0,  1,  1), lat, lat)] = hopping( 8 )
    sys[kwant.builder.HoppingKind((0,  1, -1), lat, lat)] = hopping( 9 )
    #### Define the left lead. ####
    lead = kwant.Builder(kwant.TranslationalSymmetry((-a, 0,0)))
    lead[( lat(0, j, k ) for j in range(W) for k in range(H) )] = \
        ( M0 - 2 * M1 - 4 * M2 ) * sigma_z
    # hoppings in x-direction
    lead[kwant.builder.HoppingKind((1, 0, 0), lat, lat)] = \
        M2 * sigma_z - s * 1j * v0 / 2 * sigma_x
    # hoppings in y-directions
    lead[kwant.builder.HoppingKind((0, 1, 0), lat, lat)] = \
        M2 * sigma_z + 1j * v0 / 2 * sigma_y
    # hoppings in z-directions
    lead[kwant.builder.HoppingKind((0, 0, 1), lat, lat)] = \
        M1 * sigma_z

    #### Attach the leads and return the finalized system. ####
    sys.attach_lead(lead)
    sys.attach_lead(lead.reversed())

    return sys

def plot_conductance(Ax):
    # Compute conductance

    data1 = []
    data2 = []

    for A0 in Ax:

        s =  1

        sys = make_system( s, A0  )

        sys = sys.finalized()

        smatrix = kwant.smatrix(sys, energy = Ef)

        data1.append( smatrix.transmission(1, 0) )

        s =  -1

        sys = make_system( s, A0  )

        sys = sys.finalized()

        smatrix = kwant.smatrix(sys, energy = Ef)

        data2.append(smatrix.transmission(1, 0) )

        print( A0 )

    pyplot.figure()
    pyplot.plot(Ax, data1,'k-', Ax, data2,'r--')
    pyplot.xlabel("kA")
    pyplot.ylabel("conductance [e^2/h]")
    pyplot.show()

    return data1, data2

def main():

    syst = make_system( )

    # Check that the system looks as intended.
    kwant.plot(syst)

    # Finalize the system.
    syst = syst.finalized()

    plot_conductance( Ax = np.linspace( 0.0, 2, 200) )

# Call the main function if the script gets executed (as opposed to imported).
# See <http://docs.python.org/library/__main__.html>.
if __name__ == '__main__':
    main()

\end{verbatim}

\textbf{2. Realization of the model with light confinement}

\begin{verbatim}
#!/usr/bin/env python3
# -*- coding: utf-8 -*-
"""
@author: Ming-Xun Deng
"""
import kwant

import numpy as np

import tinyarray

from matplotlib import pyplot

from scipy import special

# define the pauli matrix

sigma_0=tinyarray.array([[1,0],[0,1]])

sigma_x=tinyarray.array([[0,1],[1,0]])

sigma_y=tinyarray.array([[0,-1j],[1j,0]])

sigma_z=tinyarray.array([[1,0],[0,-1]])

## system parameters

a = 1

yita = 1

theta = 0

phi = np.pi / 2

omega = 1

Ef = 0.05

v0 = 0.24598

M0 = - 0.08686

M1 = - 0.106424

M2 = - 0.103610

Lx = 500

Gauss = 0 #  0 for plane beam and > 0 for Gaussian light beam

rho_x = np.sqrt( np.cos(theta)**2 * np.cos(phi)**2 + np.sin(phi)**2 )

rho_y = np.sqrt( np.cos(theta)**2 * np.sin(phi)**2 + np.cos(phi)**2 )

rho_z = np.sin(theta)

sin_ax = np.sin(phi) / rho_x

sin_ay = np.cos(phi) / rho_y

sin_xy = np.cos(theta) / ( rho_x * rho_y )

def make_system( s=1, A0=0, ky=0, kz=0 ):

    lat = kwant.lattice.chain( a )

    sys = kwant.Builder( )

    for x in range( Lx ):

        if Gauss > 0:

            kA = A0 * np.exp( - ( x - Lx / 2 ) * 2 / Gauss ** 2 )
        else:
            kA = A0

        J_0x = special.jv(0, kA * rho_x )

        J_0y = special.jv(0, kA * rho_y )

        J_0z = special.jv(0, kA * rho_z )

        J_1x = special.jv(1, kA * rho_x )

        J_1y = special.jv(1, kA * rho_y )

        J_1z = special.jv(1, kA * rho_z )

        lam = yita * 2 * v0 ** 2 * J_1x * J_1y * sin_xy  / omega

        v1_x = yita * 4 * M1 * v0 * J_1x * J_1z * sin_ax  / omega

        v1_y = yita * 4 * M1 * v0  * J_1y * J_1z * sin_ay / omega

        v2 = yita * 4 * M2 * v0 * J_1x * J_1y * sin_xy  / omega

        ux = v1_y * np.sin( kz * a ) * sigma_x

        uy = ( (J_0y * v0 - s * v1_x) * np.sin( ky * a )  \
            + s * v1_x * np.sin( kz * a ) ) * sigma_y

        uz = ( M0 - 2 * M1 - 4 * M2 + 2 * J_0y * M2 * np.cos( ky * a ) \
            + 2 * J_0z * M1 * np.cos( kz * a )  - s * lam ) * sigma_z

        sys[ lat(x) ] = ux + uy + uz

        if x > 0 :
            sys[ lat(x-1), lat(x) ] = J_0x * M2 * sigma_z  \
                - 1j / 2 * ( s * J_0x * v0 - v2 ) * sigma_x


    sym_lead = kwant.TranslationalSymmetry([-a])

    lead = kwant.builder.Builder(sym_lead)

    hy = v0 * np.sin( ky * a ) * sigma_y

    hz = ( M0 - 2 * M1 - 4 * M2 + 2 * M2 * np.cos( ky * a )  \
        + 2 * M1 * np.cos( kz * a )  ) * sigma_z

    lead[ lat(0) ] = hy + hz

    lead[ lat(-1),lat(0) ]= M2 * sigma_z - 1j / 2 * s * v0 * sigma_x

    sys.attach_lead(lead)
    sys.attach_lead(lead.reversed())

    return sys

def plot_conductance(Ax, ksy, ksz ):

    data1 = []

    data2 = []

    for A0 in Ax:

        # Compute conductance
        spin_up = np.zeros( ( ksz.shape[0], ksz.shape[0] ) )

        spin_dw = np.zeros( ( ksz.shape[0], ksz.shape[0] ) )

        for m in range( len(ksz) ):

            kz = ksz[m]

            for n in range( len( ksy ) ):

                ky = ksy[n]

                s =  1

                sys = make_system( s, A0, ky, kz )

                sys = sys.finalized()

                smatrix = kwant.smatrix(sys, energy = Ef)

                spin_up[ m, n ] = smatrix.transmission(1, 0)

                s =  -1

                sys = make_system( s, A0, ky, kz )

                sys = sys.finalized()

                smatrix = kwant.smatrix(sys, energy = Ef)

                spin_dw[ m, n ] = smatrix.transmission(1, 0)

            print( kz )

        data1.append( spin_up.sum(0).sum(0) * ( ksz[1]-ksz[0] )*( ksy[1]-ksy[0] ) )

        data2.append( spin_dw.sum(0).sum(0) * ( ksz[1]-ksz[0] )*( ksy[1]-ksy[0] ) )
    pyplot.figure()
    pyplot.plot(Ax, data1,'k-', Ax, data2,'r-')
    pyplot.xlabel("kz")
    pyplot.ylabel("conductance [N/4pi^2 * e^2/h]")
    pyplot.show()

    return spin_up, spin_dw

def main():

    sys = make_system( )

    # Check that the system looks as intended.
    kwant.plot(sys)
    plot_conductance( Ax = np.linspace( 0, 2, 50 ), ksy = \
        np.linspace( -0.3, 0.3, 500 ), ksz = np.linspace( -1.3, 1.3, 500 ) )

# Call the main function if the script gets executed (as opposed to imported).
# See <http://docs.python.org/library/__main__.html>.
if __name__ == '__main__':
    main()

\end{verbatim}

\textbf{3. Realization of the model without light confinement}

\begin{verbatim}
#!/usr/bin/env python3
# -*- coding: utf-8 -*-
"""
@author: Ming-Xun Deng
"""
import kwant

import numpy as np

import tinyarray

from matplotlib import pyplot

from scipy import special

# define the pauli matrix

sigma_0=tinyarray.array([[1,0],[0,1]])

sigma_x=tinyarray.array([[0,1],[1,0]])

sigma_y=tinyarray.array([[0,-1j],[1j,0]])

sigma_z=tinyarray.array([[1,0],[0,-1]])

## system parameters

a = 1

t = 0.1

mu = 0.2

yita = 1

theta = 0

phi = np.pi / 2

omega = 1

Ef = 0.05

v0 = 0.24598

M0 = - 0.08686

M1 = - 0.106424

M2 = - 0.103610

Lx = 500

Gauss = 0  # 0 plane beam and > 0 for Gaussian light beam

rho_x = np.sqrt( np.cos(theta)**2 * np.cos(phi)**2 + np.sin(phi)**2 )

rho_y = np.sqrt( np.cos(theta)**2 * np.sin(phi)**2 + np.cos(phi)**2 )

rho_z = np.sin(theta)

sin_ax = np.sin(phi) / rho_x

sin_ay = np.cos(phi) / rho_y

sin_xy = np.cos(theta) / ( rho_x * rho_y )

def make_system( s=1, A0=0, ky=0, kz=0 ):

    lat = kwant.lattice.chain( a )

    sys = kwant.Builder( )

    for x in range( Lx ):

        if Gauss > 0:

            kA = A0 * np.exp( - ( x - Lx / 2 ) * 2 / Gauss ** 2 )
        else:
            kA = A0

        J_0x = special.jv(0, kA * rho_x )

        J_0y = special.jv(0, kA * rho_y )

        J_0z = special.jv(0, kA * rho_z )

        J_1x = special.jv(1, kA * rho_x )

        J_1y = special.jv(1, kA * rho_y )

        J_1z = special.jv(1, kA * rho_z )

        lam = yita * 2 * v0 ** 2 * J_1x * J_1y * sin_xy  / omega

        v1_x = yita * 4 * M1 * v0 * J_1x * J_1z * sin_ax  / omega

        v1_y = yita * 4 * M1 * v0  * J_1y * J_1z * sin_ay / omega

        v2 = yita * 4 * M2 * v0 * J_1x * J_1y * sin_xy  / omega

        ux = v1_y * np.sin( kz * a ) * sigma_x

        uy = ( (J_0y * v0 - s * v1_x) * np.sin( ky * a )  \
            + s * v1_x * np.sin( kz * a ) ) * sigma_y

        uz = ( M0 - 2 * M1 - 4 * M2 + 2 * J_0y * M2 * np.cos( ky * a ) \
            + 2 * J_0z * M1 * np.cos( kz * a )  - s * lam ) * sigma_z

        sys[ lat(x) ] = ux + uy + uz

        if x > 0 :
            sys[ lat(x-1), lat(x) ] = J_0x * M2 * sigma_z  \
                - 1j / 2 * ( s * J_0x * v0 - v2 ) * sigma_x

    sym_lead = kwant.TranslationalSymmetry([-a])

    lead = kwant.builder.Builder(sym_lead)

    lead[ lat(0) ] = ( - mu + 6 * t - 2 * t * J_0y * np.cos( ky * a ) \
        - 2 * t * J_0z * np.cos( kz * a ) ) * sigma_0

    lead[ lat(-1), lat(0) ] = - t * J_0y * sigma_0

    sys.attach_lead(lead)
    sys.attach_lead( lead.reversed() )

    return sys

def plot_conductance(Ax, ksy, ksz ):

    data1 = []

    data2 = []

    for A0 in Ax:

        print( A0 )

        # Compute conductance
        spin_up = np.zeros( ( ksz.shape[0], ksz.shape[0] ) )

        spin_dw = np.zeros( ( ksz.shape[0], ksz.shape[0] ) )

        for m in range( len(ksz) ):

            kz = ksz[m]

            for n in range( len( ksy ) ):

                ky = ksy[n]

                s =  1

                sys = make_system( s, A0, ky, kz )

                sys = sys.finalized()

                smatrix = kwant.smatrix(sys, energy = Ef)

                spin_up[ m, n ] = smatrix.transmission(1, 0)

                s =  -1

                sys = make_system( s, A0, ky, kz )

                sys = sys.finalized()

                smatrix = kwant.smatrix(sys, energy = Ef)

                spin_dw[ m, n ] = smatrix.transmission(1, 0)

            print( kz )

        data1.append( spin_up.sum(0).sum(0) * ( ksz[1]-ksz[0] )*( ksy[1]-ksy[0] ) )

        data2.append( spin_dw.sum(0).sum(0) * ( ksz[1]-ksz[0] )*( ksy[1]-ksy[0] ) )

    pyplot.figure()
    pyplot.plot(Ax, data1,'k-', Ax, data2,'r-')
    pyplot.xlabel("kA")
    pyplot.ylabel("conductance [N/4pi^2 *e^2/h]")
    pyplot.show()

    return spin_up, spin_dw

def main():

    sys = make_system( )

    # Check that the system looks as intended.
    kwant.plot(sys)
    plot_conductance( Ax = np.linspace( 0, 2, 50 ), ksy = \
        np.linspace( -0.3, 0.3, 500 ), ksz = np.linspace( -1.3, 1.3, 500 ) )

# Call the main function if the script gets executed (as opposed to imported).
# See <http://docs.python.org/library/__main__.html>.
if __name__ == '__main__':
    main()

\end{verbatim}